# Influence of Inter-Pulse Delay and Geometric Constraints on Damage and Optical Characteristics in thin Metal Targets Irradiated by Double Ultrashort Laser Pulses

George D. Tsibidis[1,a]

[1]*Institute of Electronic Structure and Laser (IESL), Foundation for Research and Technology (FORTH), Vassilika Vouton, 70013, Heraklion, Crete, Greece*

[a] Authors to whom correspondence should be addressed: tsibidis@iesl.forth.gr

**ABSTRACT**

Femtosecond pulsed laser systems constitute powerful tools for the high-precision structuring of materials at micro/nano-scale resolutions. A critical parameter influencing the efficacy of ultrafast laser-material interactions is the laser-induced damage threshold (LIDT), which is defined as the minimum laser fluence required to induce irreversible modification to the material surface. While extensive studies have addressed single-pulse damage mechanisms, the response of thin metallic films to double-pulse femtosecond irradiation, particularly when the film thickness is of the order of the optical penetration depth, remains, generally, unexplored. In this work, we present a rigorous theoretical investigation into the spatiotemporal evolution of energy deposition, thermalization processes and optical parameter changes under double-pulse excitation conditions. The analysis considers key parameters including the inter-pulse delay and the film thickness to evaluate their influence on the LIDT for a range of technologically relevant metals: Au, Ag, Cu, Al, Ni, Ti, Cr, Pt, W, Mo, and Stainless Steel (100Cr6). The computational framework integrates a multiscale model combining laser energy absorption, non-equilibrium electron dynamics, electron–phonon coupling and subsequent thermal diffusion. A comparative analysis highlights the potential of controlled double-pulse irradiation schemes to manipulate energy coupling efficiency, improve the spatial selectivity of laser-induced modifications and compile a comprehensive LIDT database for commonly used industrial materials. The approach is aimed to provide a robust foundation for the design and optimization of advanced laser micromachining and nanofabrication protocols across a broad spectrum of metallic systems.

## 1. Introduction

Ultrashort pulsed lasers have significantly advanced precision material processing by allowing highly localized energy delivery with minimal thermal impact. The capacity to achieve precise control over energy deposition makes ultrafast lasers ideal tools for fabricating micro/nano-scale features for a broad range of materials, including metals, dielectrics, semiconductors and polymers [1]. Among the various ultrashort laser strategies which were developed, double-pulse irradiation, in which two laser pulses are temporally delayed on the order of femtoseconds to picoseconds, has emerged as a promising approach to improve control over laser-material interactions. By tailoring the inter-pulse delay, the modulation of the cumulative effects of the pulses on electron excitation, energy absorption and lattice heating can be controlled. Numerous experimental and theoretical studies have demonstrated that double-pulse schemes can lead to either enhancement or suppression of material removal, depending on the delay, fluence, and intrinsic properties of the target material [2-14].

On the other hand, a fundamental parameter in any laser-based material processing scheme is the laser-induced damage threshold (LIDT), the minimum fluence required to induce irreversible changes to the target (i.e. melting). An accurate prediction and control of the LIDT are essential for both optimizing the processing efficiency and ensuring structural integrity. In previous reports, it has been shown that the geometry of the target, especially in thin films or nanostructured layers where the material thickness is comparable to the optical penetration depth, introduces additional complexity in the evaluation of LIDT [15-22]. In those cases, thermal confinement appear to be significant which influences the damage mechanisms compared to bulk materials [15]. Furthermore, it has been shown that the optical and thermophysical properties of metals including reflectivity, absorption coefficient, electron-phonon coupling strength, and thermal conductivity, manifestly play a pivotal role in determining how energy is absorbed, distributed, and dissipated during and after laser irradiation [23-26].

Although the majority of the studies have focused on determining the LIDT for single-pulse excitation [15-22], the understanding of how this threshold is modified under double-pulse excitation, particularly in metallic systems, still remains unexplored. Despite advances in laser processing technologies, a systematic and comparative study of how the material thickness and the temporal delay between pulses collectively influence the damage thresholds under strong excitation across a wide range of metals is still insufficient.

To bridge the above knowledge gap, in this work, we present a comprehensive theoretical investigation of the influence of material properties, inter-pulse delay and material thickness on the laser-induced damage behaviour of metallic targets exposed to temporarily delayed ultrashort laser pulses. A selection of commonly used and industrially relevant metals including Au, Ag, Cu, Al, Ni, Ti, Cr, Pt, W, Mo and 100Cr6, is investigated to highlight material-specific trends. A theoretical approach based on the use of the two-temperature model (TTM) is followed to simulate the ultrafast dynamics of electron and lattice subsystems, incorporating energy absorption, electron-phonon relaxation, and thermal diffusion [15, 25, 27]. A systematic investigation is performed by varying the inter-pulse delay, film thickness in order to identify regimes where laser-induced damage is either enhanced or reduced while the optical parameter correlation with the film thickness and the pulse delay are also evaluated.

## 2. Theoretical Model

The primary objective of this investigation is to elucidate the influence of intrinsic material parameters and external laser parameters in modulating the LIDT under double-pulse irradiation. To model ultrafast dynamics and thermal effects in a metal-dielectric bilayered structure, a one-dimensional Two Temperature Model (1D-TTM) [27] through a set of coupled rate equations (Eqs.1) is employed. The model considers femtosecond laser pulses of laser wavelength $\lambda_L = 1026$ nm and pulse duration $\tau_p = 170$ fs, assuming lateral uniformity due to the large laser spot size relative to film thickness (i.e. radius ~15 μm) [15]. This justifies a 1D approach along the energy propagation direction. The presence of the dielectric substrate is incorporated in the model (Eqs.1). Fused Silica is considered as a substrate although other types of materials could also be used [28]

$$
\begin{aligned}
C_e^{(m)} \frac{\partial T_e^{(m)}}{\partial t} &= \frac{\partial}{\partial z}\left(k_e^{(m)} \frac{\partial T_e^{(m)}}{\partial z}\right) - G_{eL}^{(m)}\left(T_e^{(m)} - T_L^{(m)}\right) + S^{(m)} \\
C_L^{(m)} \frac{\partial T_L^{(m)}}{\partial t} &= \frac{\partial}{\partial z}\left(k_L^{(m)} \frac{\partial T_L^{(m)}}{\partial z}\right) + G_{eL}^{(m)}\left(T_e^{(m)} - T_L^{(m)}\right) \\
C_L^{(S)} \frac{\partial T_L^{(S)}}{\partial t} &= \frac{\partial}{\partial z}\left(k_L^{(S)} \frac{\partial T_L^{(S)}}{\partial z}\right) + S^{(S)}
\end{aligned}
\qquad (1)
$$

where the subscript '$m$' (or '$S$') indicates the thin film (or substrate). In Eqs.1, $T_e^{(m)}$ and $T_L^{(m)}$ corresponds to the electron and lattice temperatures, respectively, of the metal film. The thermophysical properties of the metal such as the electron $C_e^{(m)}$ and lattice $C_L^{(m)}$ volumetric heat capacities, electron $k_e^{(m)}$ $\left(= k_{e0}^{(m)} \frac{B_e T_e^{(m)}}{A_e\left(T_e^{(m)}\right)^2 + B_e T_L^{(m)}}\right)$ heat conductivity, the electron-phonon coupling strengths $G_{eL}^{(m)}$, $A_e$, $B_e$ and other model thermophysical parameters in Eqs.1 are listed in Table 1. It is emphasized that heat conduction in metals is primarily governed by electrons, meaning that the thermal conductivity of the lattice is significantly smaller than that of the electron system. To reflect this difference, $k_L^{(m)}$ is approximated as a small fraction of the electron heat conductivity, such as ($k_L^{(m)}$=0.01$k_e^{(m)}$) a convention adopted in previous studies [29, 30].

The quantity $S^{(m)}$ in Eqs.1 represents the laser-induced source term that delivers energy to the metal surface and it is sufficient to excite carriers within the thin film. Since this study focuses on the effects of optically excited *thin* films, several key factors must be considered: (i) part of the laser energy is absorbed by the film while some is transmitted into the substrate, (ii) reflectivity and transmissivity are affected by multiple reflections at the air/metal and metal/substrate interfaces, and (iii) the transmitted energy into the substrate is insufficient to generate excited carriers. As a result, the third equation in Eqs.1 includes a term $S^{(S)}$ that only heats the substrate lattice; $T_L^{(S)}$, $C_L^{(S)}$, $k_L^{(S)}$ stands for the substrate temperature, volumetric heat capacity and heat conductivity, respectively. The expression for the source term $S^{(m)}$ which is used to describe the excitation of a metallic surface of thickness $d$ is provided from the following [16]

$$S^{(m)} = \frac{(1-R-T)\sqrt{4\log(2)}F}{2\sqrt{\pi}\tau_p(\alpha^{-1}+L_b)} \frac{1}{(1-\exp(-d/(\alpha^{-1}+L_b)))} \left[\exp\left(-4\log(2)\left(\frac{t-3t_p}{t_p}\right)^2\right) + \exp\left(-4\log(2)\left(\frac{t-3t_p-t_d}{t_p}\right)^2\right)\right] \quad (2)$$

where $R$ and $T$ stand for the reflectivity and transmissivity, respectively, $L_b$ corresponds to the ballistic length, $\alpha$ is the absorption coefficient that is wavelength dependent, and $F$ is the peak fluence of the laser beam. The ballistic transport is also included in the expression as it has been demonstrated that it plays significant role in the response of the material [15, 16]. Finally, $t_d$ in Eq.2 represents the time delay between the two pulses comprising the double pulse.

The calculation of $R$ and $T$ and the absorbance $A=1-R-T$ are derived through the use of the multiple reflection theory via the employment of the following expressions [15, 31] (assuming a *p*-polarised beam)

$$R = |r_{dl}|^2, \quad T = |t_{dl}|^2 \widetilde{N}_S$$

$$r_{dl} = \frac{r_{am}+r_{mS}e^{2\beta j}}{1+r_{am}r_{mS}e^{2\beta j}}, \quad t_{dl} = \frac{t_{am}t_{mS}e^{\beta j}}{1+r_{am}r_{mS}e^{2\beta j}} \quad (3)$$

$$\beta = 2\pi d/\lambda_L, r_{CD} = \frac{\widetilde{N}_D-\widetilde{N}_C}{\widetilde{N}_D+\widetilde{N}_C}, \quad t_{CD} = \frac{2\widetilde{N}_C}{\widetilde{N}_D+\widetilde{N}_C}$$

where the indices $C=a,m$ and $D=m,S$ characterise each material ('$a$', '$m$', '$S$' stand for 'air', 'metal', 'substrate', respectively). The complex refractive indices of the materials such as air, metal and substrate are denoted with $\widetilde{N}_a = 1$, $\widetilde{N}_m = Re(\widetilde{N}_m) + Im(\widetilde{N}_m)j$, $\widetilde{N}_S = Re(\widetilde{N}_s)$, respectively. Given that fused silica glass is used as a substrate material, $Re(\widetilde{N}_s)(\lambda_L = 1026\ nm) \cong 1.4501$ [32]. The approach can be used for any substrate [28]. The calculation of the dielectric function for each metal is based on the analysis by Rakic et al. via the use of Drude-Lorentz model (where both interband and intraband transitions are assumed) [33] considering a temperature dependence of the reciprocal of the electron relaxation time $\tau_e$ (i.e. $\tau_e = \left[A_e\left(T_e^{(m)}\right)^2 + B_e T_L^{(m)}\right]^{-1}$) [34]. The refractive indices of the metals investigated in this study (at 300 K) are given in Table 1.

| Parameter | Material | | | | | | | | | | |
|---|---|---|---|---|---|---|---|---|---|---|---|
| | Au | Ag | Cu | Al | Ni | Ti | Cr | Steel (100Cr6) | W | Pt | Mo |
| $\widetilde{N}_m$ (at $\lambda_L$=1026 nm) | DL [33] | DL [33] | DL [33] | DL [33] | DL [33] | DL [33] | DL [33] | DL [34] | DL [33] | DL [33] | 2.4357+ i4.1672 [35] |
| $G_{eL}^{(m)}$ [Wm$^{-3}$K$^{-1}$] | Ab-Initio [36] | Ab-Initio [36] | Ab-Initio [36] | Ab-Initio [36] | Ab-Initio [36] | Ab-Initio [36] | Ab-Initio [36] | Ab-Initio [37] | Ab-Initio [36] | Ab-Initio [36] | Ab-Initio [36] |
| $C_e^{(m)}$ [Jm$^{-3}$K$^{-1}$] | Ab-Initio [36] | Ab-Initio [36] | Ab-Initio [36] | Ab-Initio [36] | Ab-Initio [36] | Ab-Initio [36] | $A_g T_e^{(m)}$ [16] | Ab-Initio [37] | Ab-Initio [36] | Ab-Initio [36] | Ab-Initio [36] |
| $C_L^{(m)}$ [×10$^6$ Jm$^{-3}$K$^{-1}$] | 2.48 [16] | 2.5 [30] | 3.3 [16] | 2.4 [38] | 4.3 [16] | 2.35 [39] | 3.3 [16] | 3.27 [37] | 2.5 [40] | 2.8 [40] | 2.33 [40] |
| $k_{e0}^{(m)}$ [Wm$^{-1}$K$^{-1}$] | 318 [16] | 428 [30] | 401 [16] | 235 [38] | 90 [16] | 21.9 [39] | 93.9 [16] | 46.6 [37] | 173 [40] | 72 [40] | 138 [40] |
| $A_e$ [×10$^7$ s$^{-1}$K$^{-2}$] | 1.18 [30] | 0.932 [30] | 1.28 [30] | 0.376 [38] | 0.59 [30] | 1 [39] | 7.9 [29] | 0.98 [37] | 1 [39] | 1 [39] | 1 [39] |
| $B_e$ [×10$^{11}$ s-1K-1] | 1.25 [30] | 1.02 [30] | 1.23 [30] | 3.9 [38] | 1.4 [30] | 1.5 [39] | 13.4 [29] | 2.8 [37] | 1.5 [39] | 1.5 [39] | 1.5 [39] |
| $T_{melt}$ [K] | 1337 [40] | 1234 [40] | 1357 [40] | 933 [40] | 1728 [40] | 1941 [40] | 2180 [40] | 1811 [40] | 3695 [40] | 2045 [40] | 2896 [40] |

**Table 1:** Optical and thermophysical properties of materials (*DL* stands for Drude-Lorentz model); $A_g$=194 Jm$^{-3}$K$^{-1}$.

The volumetric heat capacity of fused silica glass is $C_L^{(S)}$ =1.6×10$^6$ Jm$^{-3}$K$^{-1}$ while its heat conductivity is equal to $k_L^{(S)}$=1.3 Wm$^{-1}$K$^{-1}$. An iterative Crank-Nicolson scheme based on a finite-difference method is used to solve Eqs.1-3. The physical system is considered to be in thermal equilibrium at *t*=0 and, therefore, $T_e^{(m)}(z,t=0) = T_L^{(m)}(z,t=0) =$

300 K. A thick substrate is assumed (i.e. $k_L^{(S)} \frac{\partial T_L^{(S)}}{\partial z} = 0$) while adiabatic conditions are applied on the surface of the metal surface (i.e. $k_e^{(m)} \frac{\partial T_e^{(m)}}{\partial z} = 0$). To solve numerically Eqs.1-3, the following boundary conditions are considered on the interface between the top layer and the substrate: $k_L^{(m)} \frac{\partial T_L^{(m)}}{\partial z} = k_L^{(S)} \frac{\partial T_L^{(S)}}{\partial z}, k_e^{(m)} \frac{\partial T_e^{(m)}}{\partial z} = 0, T_L^{(m)} = T_L^{(S)}$. The iterative approach towards evaluating LIDT is described in detail in the Supplemntary Material.

## 3. Results and Analysis

### A. Damage threshold

The physics-based theoretical model presented in the previous section is aimed to describe the ultrafast dynamics, optical behavior and thermal response of the electron and lattice subsystems in metals. It focuses on a detailed exploration of the role of both the inter-pulse delay between laser sub-pulses and the material thickness. By investigating the effects of geometrical constraints and the absorbed energy associated with the temporal spacing of the laser pulses, the model highlights how the physical properties of metals impact the laser-induced damage threshold (LIDT) and their optical characteristics. The simulations aim to explore how the thermal and optical behavior of irradiated materials depend on the type of metal. In particular, a detailed analysis is conducted to elucidate how the intrinsic properties and transient changes in the thermophysical parameters of metals influence their thermal response. This analysis is performed across eleven representative metals and the results are organized based on the characteristics of the metals originated from their electronic structure: (1) transition metals with nearly filled *d*-bands: nickel, platinum, chromium, and molybdenum; (2) aluminum; (3) noble metals: gold, silver, copper; (4) transition metals with less than half-filled *d*-bands: tungsten, and titanium; (5) an alloy: stainless steel.

Figures 1-11 illustrate results based on the analysis of the induced thermal effects. More specifically, Figures 1–11 (a, c, d) show filled contour plots of simulated data displaying the damage threshold variation with pulse separation and material thickness (a), depicting how these results scale with single pulses (LIDT-SP) as a function of the material thickness (c) and illustrating the deviation from the bulk sample predictions (d). The areas between contour lines are colour-shaded to represent different data ranges. The number of contour levels (or their exact values) are adjusted to control the level of detail in how the data range is divided. Furthermore, the electron and lattice temperature evolution are displayed in Figures 1-11 (b) for two different film thickness values, *d*=20 nm (which is comparable to the optical penetration depth) and *d*=210 nm (approaching bulk-like behaviour). The transient variation of the temperatures is calculated at $F=DT_d^{SP}$, where $DT_d^{SP}$ denotes the damage threshold under single-pulse irradiation for a metal film of thickness *d*. It is noted that simulation results are shown for all material for *d* ranging from 10 nm to 310 nm and $t_d$ from 0 fs (single pulse) to 25 ps. Thus, for film thicknesses much smaller than the laser spot radius, heat transport is predominantly along the depth direction. Thus, a 1D description is sufficient to capture the dominant physics relevant to damage threshold determination. By contrast, for thicker films or smaller spot sizes, lateral diffusion may become significant and an extension to 2D or 3D modeling approach is required, however, this beyond the scope of the current work. Material-specific trends highlight the crucial role of intrinsic thermophysical properties in determining the damage behavior. The simulations reveal that the LIDT of metallic films is strongly determined by the interplay between the material properties, geometrical confinement and temporal separation of double ultrashort laser pulses. In principle, the behaviour of each material under double ultrashort laser pulse irradiation is dictated by its thermal and optical properties, specifically its electron thermal conductivity and electron-phonon coupling strength. A fundamental observation across all materials is that the LIDT decreases significantly as the film thickness is reduced (from 100 nm down to 10 nm). This effect is pronounced, leading to a reduction in the LIDT by approximately 73% to 88% relative to the bulk material response (Figures 1-11d); this is due to the thermal confinement arises from limitations on electron diffusion at smaller thicknesses. Furthermore, the results presented in the Supplementary Material demonstrate how material thickness and pulse separation affect the amount of the absorbed energy, which is a key factor that influences the resulting thermal effects and the predicted LIDT. Below, the dependence of LIDT as a function of $t_d$ and *d* is discussed in detail for each material. The filled contour plots presented in Figures 1-11 were generated from discretely sampled simulation data for pulse separation and film thickness. Minor irregularities in the contour lines result from interpolation between these discrete grid points and do not reflect physical discontinuities. Grid convergence tests confirmed that a refinement of the spatial or temporal resolution does not affect the calculated damage thresholds or the qualitative behaviour. A larger number of contour lines would make more difficult the visibility while a finer grid (i.e. more spatial and temporal points and smaller step) would delay the acquisition of the map while the behaviour would not alter from the current used. While more contour lines may reduce clarity, refining the grid with additional spatial and temporal points minimally affects acquisition time and does not alter the observed behavior.

## 1. Nickel

Nickel demonstrates one of the most pronounced $t_d$-dependent responses among the explored metals in this study. Simulation results demonstrate an initial drop of the Damage Threshold (termed as DT in Figures 1 (a,c)) for short inter-pulse delays ($t_d<\tau_{e-ph}\approx$8-10 ps [21, 24] and Supplementary Material) and relatively large thicknesses followed by a minimum value occurring for pulse separations close to the electron-phonon relaxation time $\tau_{e-ph}$, before an increase of LIDT relaxing to the single pulse LIDT value. A projection of the LIDT results on the maximum lattice temperature which is attained (i.e. that exhibits an inverse relationship with LIDT) suggest that it increases initially, then, at $t_d\approx \tau_{e-ph}$ it reaches a maximum value before a subsequent slow decrease. An analysis of the ultrafast phenomena and thermal effects as well as the thermophysical properties of the irradiated solid are required to interpret this non-monotonic behaviour of LIDT and maximum temperature of the surface of the material. According to Table I, Ni is characterized by a strong electron-phonon coupling strength and moderate thermal conductivity and therefore, it exhibits a distinct ultrafast energy relaxation pathway under double-pulse femtosecond excitation. The following physical processes occur at the three timescales dictated by the pulse separation range compared to $\tau_{e-ph}$:

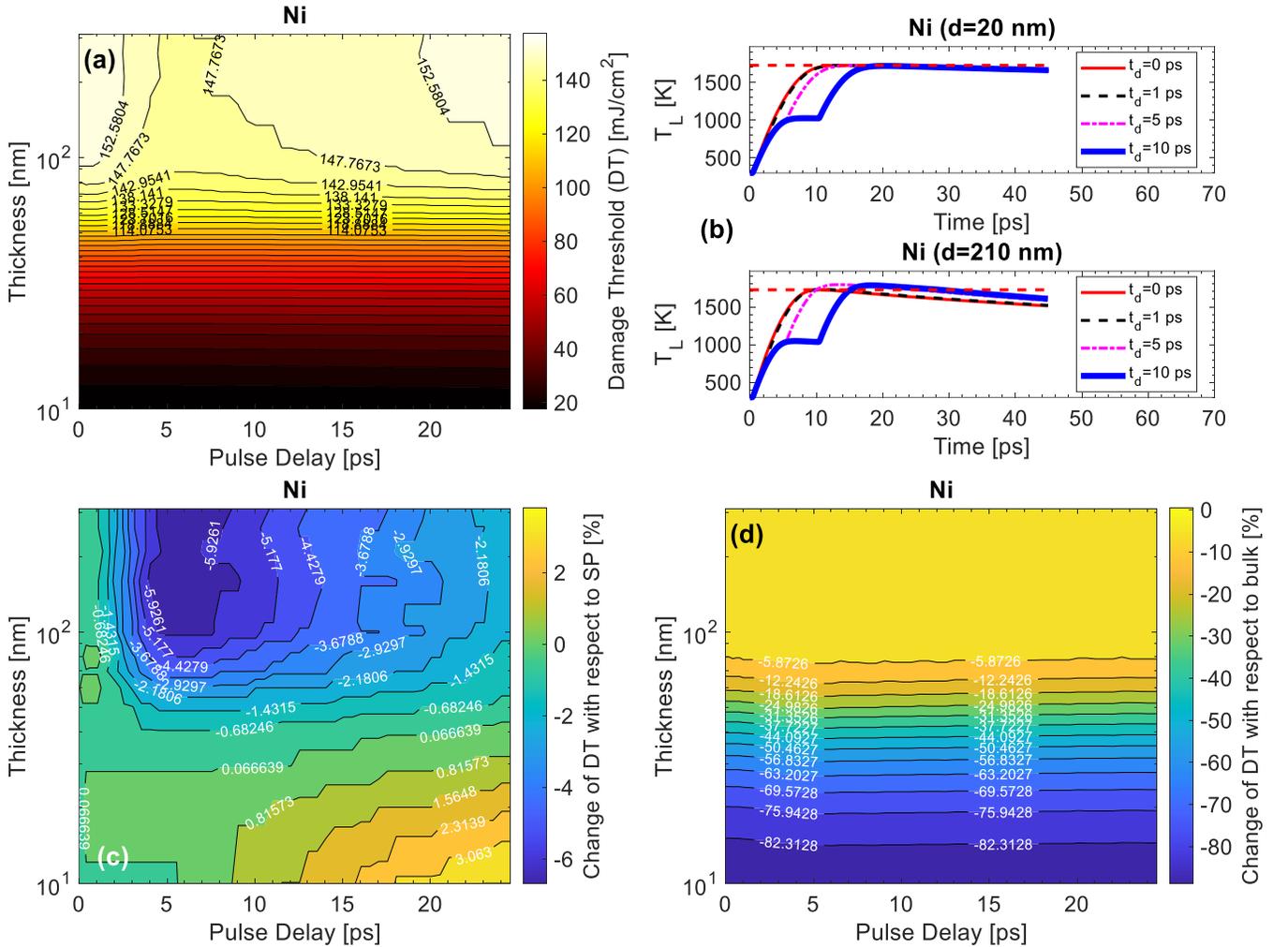

**Figure 1:** Results for Ni at different pulse delays and material thicknesses: (a) Damage Threshold (*DT*), (b) Lattice temperature evolution for *d*=20 nm and *d*=210 nm for $F=DT_d^{SP}$ at different time delays $t_d$ (the horizontal *red* dashed line represents the melting point temperature and is shown as a guide to the eye), (c) Percentage change in *DT* relative to single-pulse exposure, and (d) Percentage change in *DT* relative to bulk material response.

For thick films (*d*=210 nm) and for short inter-pulse delays ($t_d<\tau_{e-ph}$), the second component of the double laser pulse interacts with a non-equilibrated electronic subsystem while the electron temperature remains elevated as a result of the excitation from the first component of the laser pulse (see Supplementary Material). Due to the strong electron-phonon coupling (EPC), a rapid energy transfer from the hot electron population to the lattice dictates the relaxation process. As a result, the lattice temperature increases abruptly, leading to an intense localized heating. On the other hand, although a portion of the electronic energy has already been transferred to the lattice, a complete thermal equilibration has not yet been accomplished as this occurs at a timescale defined by $\tau_{e\text{-}ph}$. This temporal overlap enhances energy absorption due to an observed increased absorptivity (see Supplementary Material). Thus, the laser energy deposition is amplified, and LIDT is reduced compared to single-pulse irradiation.

For pulse separation of the order of $\tau_{e\text{-}ph}$ ($t_d \approx \tau_{e\text{-}ph}$), a thermal equilibration between the electron and lattice subsystems occurs partially before the application of the second pulse. The electronic system remains moderately hot, while the lattice has already absorbed a significant portion of the energy deposited by the first pulse. Due to the relatively low thermal conductivity of Ni, the accumulated heat remains spatially localized maintaining enhanced local heating that further elevates the lattice temperature and consequently lowers the LIDT. However, as the system begins to thermally recover, the temperatures of the electron and lattice subsystems approach equilibrium; thus, the LIDT begins to increase compared to shorter delay times, as partial cooling between pulses (i.e. lattice temperature decrease) occurs.

Finally, for longer inter-pulse spacing ($t_d > \tau_{e\text{-}ph}$), a full thermal equilibration between the electron and lattice subsystems is completed followed by partial cooling (due to diffusion) of the irradiated region. In these conditions, the second pulse interacts with an approximately equilibrated surface, and the system demonstrates negligibly enhanced thermal effects. At these delays, thermal diffusion effectively reduces the localized temperature gradients, and both the electron and lattice temperatures return close to conditions before the second pulse heats the solid (see Supplementary Material). As a result, the LIDT, further increases, gradually approaching that of single-pulse irradiation. Thus, the material response resembles that of two temporally independent single-pulse interactions, and the enhancement in laser absorption observed at shorter delays diminishes.

The trend described above is supported by simulations of lattice temperature evolution, which show maximum temperatures of $T_L^{max}$=1728, 1731, 1793, 1784 K, for $t_d$ = 0, 1, 5, 10 ps, respectively (lower panel of Figure 1b).

By contrast, thin films ($d$=20 nm) exhibit a distinctly different behaviour at short pulse separations (upper panel of Figure 1b and Figures 1(a,c)), where the LIDT as a function of pulse delay differs from that observed for a single pulse. More specifically, an increase in the LIDT is expected at increasing pulse separation, as the small thickness of the material restricts electron diffusion, rendering the contribution of electron thermal conductivity negligible relative to electron-phonon coupling and the relaxation process. In other words, the heat is highly confined, and therefore, the electron and lattice subsystems equilibrate quickly and uniformly. Because of this rapid homogenization, the second pulse interacts with a relatively equilibrated system. The maximum lattice temperatures attained in the simulations for $t_d$ = 0, 1, 5, 10 ps are $T_L^{max}$=1728, 1725, 1724, 1718 K, respectively, which confirm an expected increase of the LIDT.

Overall, according to the theoretical predictions illustrated in Figure 1a, LIDT can vary from 18-158 mJ/cm$^2$ for the range of pulse separation and material thickness values investigated in this study for Ni.

2. **Platinum**

The simulations (Figures 2(a,c)) indicate a similar qualitative trend for Pt both for thin and thick films. In particular, for thick films ($d$=210 nm), an initial drop of the LIDT at $t_d < \tau_{e\text{-}ph}$ ($\tau_{e\text{-}ph} \approx$4-5 ps for Pt, see Supplementary Material) occurs followed by a minimum value for LIDT at around $\tau_{e\text{-}ph}$ before an increase of the damage threshold. Ni and Pt are characterized by both a strong EPC, however, Pt has a smaller electron heat conductivity than Ni which is expected to influence the thermal response of the material. More specifically, our simulations reveal a small drop in LIDT for platinum compared to nickel, particularly at short pulse separations comparable to $\tau_{e\text{-}ph}$. As in the case of Ni, the decrease indicates that the second pulse interacts with a target that has not yet fully relaxed from the initial excitation producing higher transient temperatures. On the other hand, the small reduction compared to Ni is due to the fact that at short delays, the localized heating is relatively small due to the reduced heat diffusion that restrains the production of thermal effects which can result in a significant decrease in LIDT.

The lower electron conductivity is, also, responsible for the differences in increase of LIDT (Figures 2(a,c)) at longer pulse delays ($t_d > \tau_{e\text{-}ph}$) between Ni and Pt. As in the case of Ni, a rise in LIDT values at increasing inter-pulse delay occurs. This behaviour corresponds to the gradual relaxation of the electron and lattice subsystems toward thermal equilibrium before the second component of the double pulse irradiates the solid. Thus, at the time when the second pulse starts to heat the material, a substantial portion of the residual heat has dissipated, thereby reducing the accumulative thermal effects (upper panel of Figure 2b) resulting to an increase of LIDT. Results for $d$=210 nm show maximum temperatures of $T_L^{max}$=2045, 2030, 2057, 1992 K, for $t_d$ = 0, 1, 3, 10 ps, respectively (lower panel of Figure 2b) that confirm the above interpretation at short and long pulse separations. Comparing the behaviour of Ni and Pt, the larger increase of LIDT observed in platinum further highlights the stronger heat confinement and slower thermal equilibration associated with its lower electron thermal conductivity.

For thinner films (*d*=20 nm), similarly to Ni, an increase in LIDT is expected at increasing pulse separation, as the small thickness of the material restricts electron diffusion. Due to the fact that the heat is significantly confined, the electron and lattice subsystems equilibrate quickly and uniformly. Our predictions for LIDT are also reflected on the decrease of the maximum lattice temperatures: for *d*=20 nm, simulations yield maximum temperatures of $T_L^{max}$=2047, 2045, 2044, 2023 K, for $t_d$ = 0, 1, 3, 10 ps, respectively (upper panel of Figure 2b). A larger increase of LIDT for Pt compared to Ni for thinner films is (Figures 2(a,c)), also, attributed to the lower heat conductivity of Platinum.

Finally, the more rapid decrease in the lattice temperature of Pt compared to Ni after reaching its peak (Figure 1b, Figure 2b) is attributed to the lower heat capacity of Pt, which allows it to cool down faster. Overall, according to the theoretical predictions illustrated in Figure 2a, LIDT can vary from 14-117 mJ/cm² for the range of pulse separation and material thickness values investigated in this study for Pt.

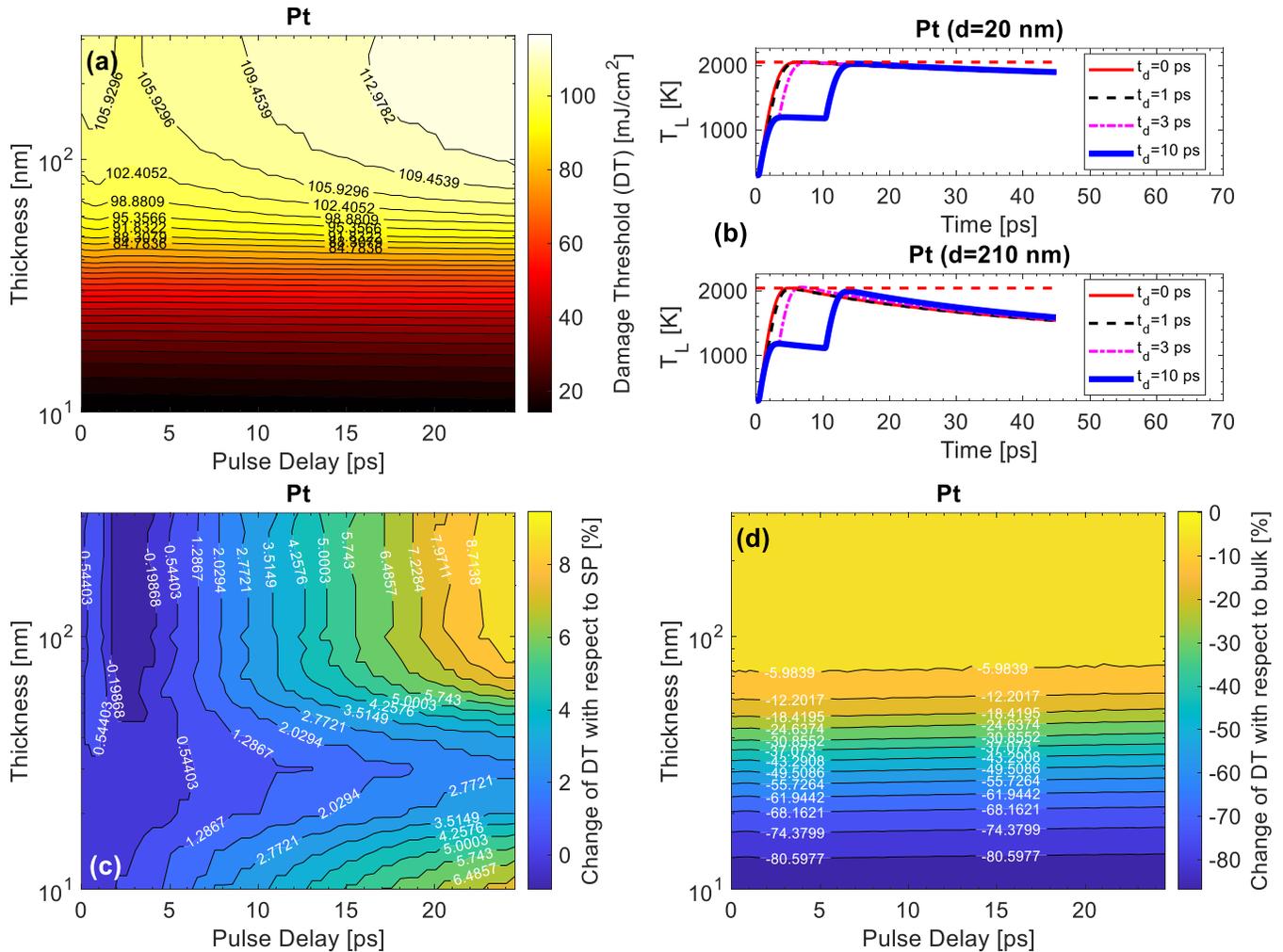

**Figure 2:** Results for Pt at different pulse delays and material thicknesses: (a) Damage Threshold (*DT*), (b) Lattice temperature evolution for *d*=20 nm and *d*=210 nm for $F=DT_d^{SP}$ at different time delays $t_d$ (the horizontal *red* dashed line represents the melting point temperature and is shown as a guide to the eye), (c) Percentage change in *DT* relative to single-pulse exposure, and (d) Percentage change in *DT* relative to bulk material response.

## 3. Chromium

Chromium is a transition metal characterized by a comparable thermal conductivity to Ni and Pt and a strong electron-phonon coupling which scales with the excitation level differently from that of the other two materials (i.e. it increases with increasing electron temperature value [41]). Simulation results illustrated in Figures 3(a,c) indicate a rise of the LIDT at increasing pulse separation. Theoretical predictions for the evolution of lattice temperature at $t_d$ = 0, 1, 5, 10 ps (for *d*=210 nm) show a decreasing trend for the maximum temperature values ($T_L^{max}$=2180, 2143, 2099, 2047 K shown in the lower panel in Figure 3b) which confirm the aforementioned pulse delay dependence of LIDT. A similar monotonic behaviour is also is demonstrated for the LIDT for thinner films (for $d$=20 nm) which is confirmed by the trend of the maximum temperatures attained at $t_d = 0, 1, 5, 10$ ps ($T_L^{max}$=2180, 2181, 2172, 2157 K shown in the upper panel in Figure 3b). A combination of the electron heat conductivity and the EPC values determine the behaviour at increasing pulse spacing: before the second pulse irradiates the material a transport of the electrons into deeper depths at increasing pulse spacing occur and therefore fewer electrons interact with the electron system. On the other hand, the coupling of these less energetic electrons with the phonon system has a smaller strength which delays the equilibration and therefore the attained maximum lattice energy is smaller. These results are reflected on the simulated $T_L^{max}$ at various pulse separations and film thicknesses (Figures 3(a,c)). It is evident that the variation of $T_L^{max}$ for smaller thicknesses is decreased due to energy confinement; on the other hand, LIDT deviation from the single-pulse case increases with both pulse delay and increasing material thickness.

Overall, according to the theoretical predictions illustrated in Figure 3a, LIDT can vary from 16-106 mJ/cm² for the range of pulse separation and material thickness values investigated in this study for Cr.

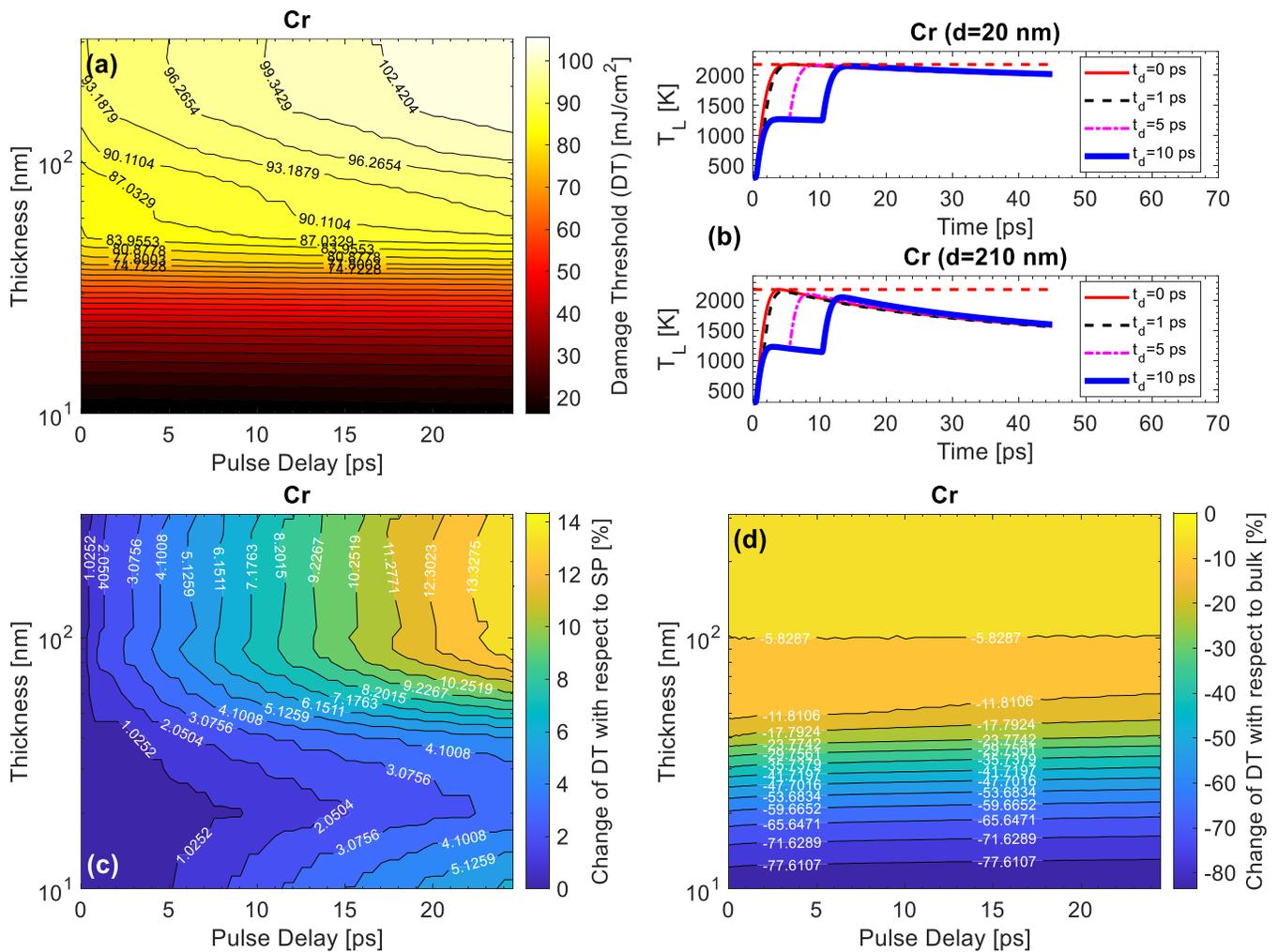

**Figure 3:** Results for Cr at different pulse delays and material thicknesses: (a) Damage Threshold (DT), (b) Lattice temperature evolution for $d$=20 nm and $d$=210 nm for $F=DT_d^{SP}$ at different time delays $t_d$ (the horizontal *red* dashed line represents the melting point temperature and is shown as a guide to the eye), (c) Percentage change in *DT* relative to single-pulse exposure, and (d) Percentage change in *DT* relative to bulk material response.

## 4. Molybdenum

Molybdenum is a transition metal of high melting point with nearly filled *d*-bands characterized by a larger electron conductivity than Cr while the EPC is, approximately, ten times smaller than that of Cr, however, it follows a similar trend at increasing electron temperature. Simulation results illustrated in Figures 4(a,c) indicate a rise of the LIDT at increasing pulse separation. Theoretical predictions for the evolution of lattice temperature $t_d = 0, 1, 5, 10$ ps (for $d$=210 nm) show a decreasing trend for the maximum temperature values ($T_L^{max}$=2896, 2816, 2688, 2581 K shown in the lower panel in Figure 4b) which confirm the aforementioned pulse delay dependence of LIDT. A similar monotonic behaviour is also demonstrated for the LIDT for thinner films (for $d$=20 nm) which is confirmed by the trend of the maximum temperatures attained ($T_L^{max}$=2896, 2870, 2834, 2802 K) illustrated in the upper panel in Figure 4b. A combination of the electron heat conductivity and the EPC values determine the behaviour at increasing pulse spacing: before the second pulse irradiates the material a transport of the electrons into deeper depths at increasing pulse spacing and therefore fewer electrons interact with the electron system. On the other hand, the coupling of these less energetic electrons with the phonon system has a smaller strength which delays the equilibration and therefore the attained maximum lattice energy is smaller. These results are reflected on the simulated $T_L^{max}$ at various pulse separations and film thicknesses (Figures 4a,c). It is evident that the variation of $T_L^{max}$ for smaller thicknesses is decreased due to energy confinement; on the other hand, LIDT deviation from the single-pulse case increases with both pulse delay and increasing material thickness. It is also noted that the smaller heat capacity, larger electron heat conductivity as well as the larger EPC compared to Cr result in a faster relaxation of the LIDT and a larger deviation from the single pulse LIDT value.

Overall, according to theoretical predictions illustrated in Figure 4a, LIDT can vary from 21-140 mJ/cm$^2$ for the range of pulse separation and material thickness values investigated in this study for Mo. Interestingly, although the electron temperature rises substantially during irradiation, efficient heat conduction quickly redistributes energy, preventing strong lattice overheating (see Supplementary Material).

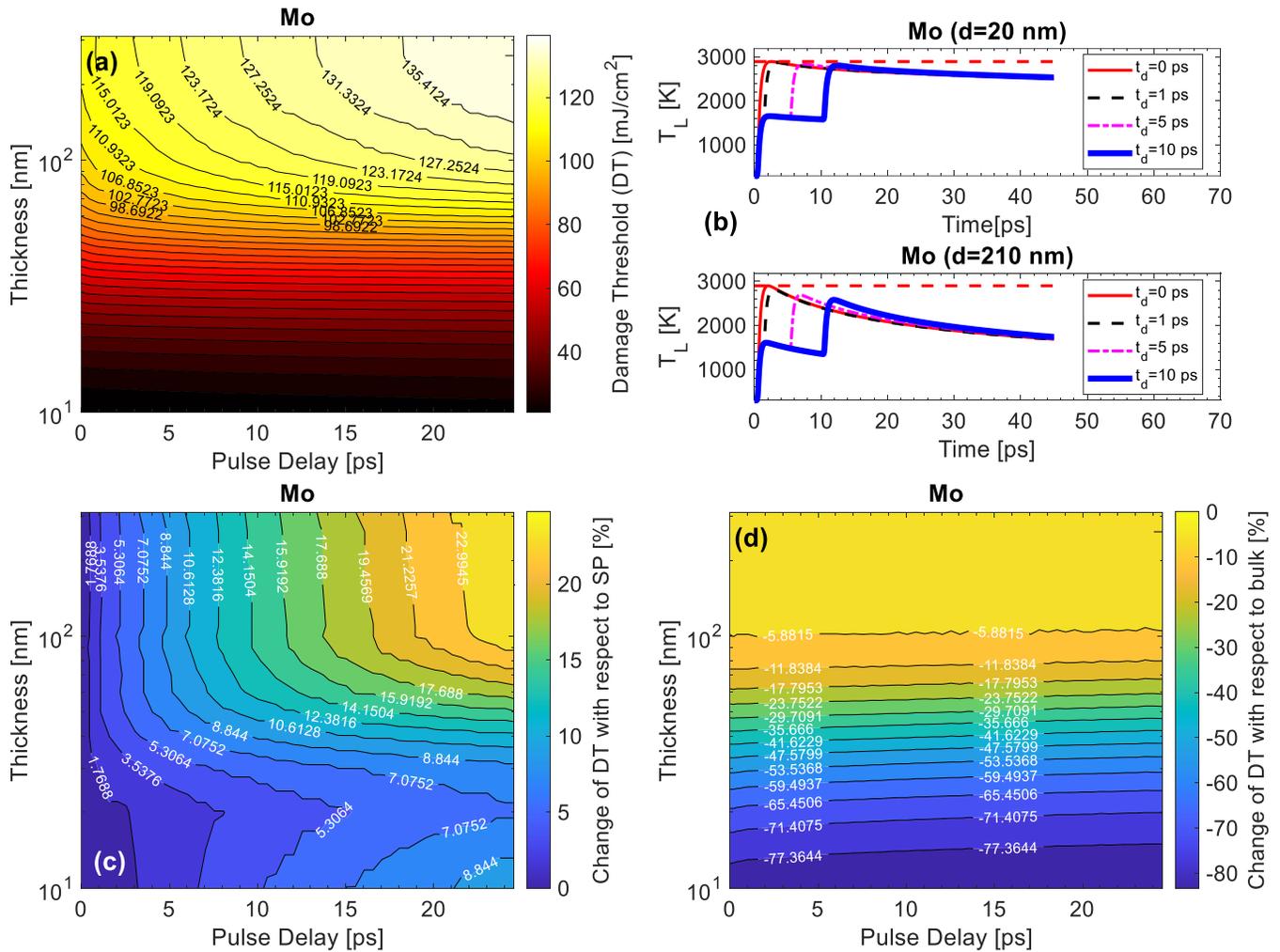

**Figure 4:** Results for Mo at different pulse delays and material thicknesses: (a) Damage Threshold (*DT*), (b) Lattice temperature evolution for *d*=20 nm and *d*=210 nm for $F=DT_d^{SP}$ at different time delays $t_d$ (the horizontal *red* dashed line represents the melting point temperature and is shown as a guide to the eye), (c) Percentage change in *DT* relative to single-pulse exposure, and (d) Percentage change in *DT* relative to bulk material response.

## 5. Aluminum

Aluminum is characterized by high electron heat conductivity and moderate electron-phonon coupling. Compared to Ni and Pt, aluminum shows a similar trend, as illustrated in Figures 5(a,c), primarily due to comparable magnitudes of EPC strength and electron diffusion (for the same reasons discussed for those metals): an initial drop following an increase of LIDT. However, its higher electron thermal conductivity enables this behavior to persist even at smaller thicknesses. The electron system (see Supplementary Material) responds similarly to that of pure transition metals: rapid excitation and fast energy transfer to the lattice. This behaviour is also reflected in the thermal response of the system. An analysis of the maximum $T_L^{max}$ with increasing pulse separation reflects the LIDT behaviour ($T_L^{max}$ exhibits the opposite behaviour to LIDT). This is confirmed by the analysis of two representative Al film thicknesses: $d$=20 nm and $d$=210 nm for which $T_L^{max}$ = 933, 943, 950, 939 K and $T_L^{max}$ = 933, 937, 941, 917 K (for the four values of $t_d$), respectively, at four pulse separations $t_d$ = 0, 0.5, 3, 10 ps. The LIDT values for the conditions investigated in this work increases with $t_d$ and it ranges between 10 mJ/cm$^2$ and 254 mJ/cm$^2$.

**Figure 5:** Results for Al at different pulse delays and material thicknesses: (a) Damage Threshold (*DT*), (b) Lattice temperature evolution for $d$=20 nm and $d$=210 nm for $F=DT_d^{SP}$ at different time delays $t_d$ (the horizontal *red* dashed line represents the melting point temperature and is shown as a guide to the eye), (c) Percentage change in *DT* relative to single-pulse exposure, and (d) Percentage change in *DT* relative to bulk material response.

## 6. Gold, Silver, Copper

The three noble metals, Au, Ag and Ag are characterized by high electron heat conductivity (slightly higher than that of Al) and low electron-phonon coupling. As a result, equilibration of the electron and lattice subsystems occurs at significantly longer times ($\tau_{e-ph}$=15-30 ps depending on $d$, see Supplementary Material) and therefore the slow transfer of energy to the lattice allows electrons to remain highly energetic for several picoseconds after the first pulse (see Figures 6b-8b and Supplementary Material). Due to the weak EPC strength, accumulation of energy is expected over long pulse separations $t_d$ in contrast with what happens for materials with faster equilibration (i.e. Ni and Pt). On the other hand, a counteracting effect arises from the high electron thermal conductivity which reduces energy localization by promoting the diffusion of highly energetic electrons away from the irradiated region. Thus, the combination of the two factors lead to high values of LIDT for the three metals (Figures 6a-8a) and remarkably larger deviations of the LIDT at various delay times relative to the single-pulse threshold (Figures 6c-8c). In particular, the LIDT ranges for the three materials are: 21-721 mJ/cm² (for Au), from 23-1176 mJ/cm² (for Ag), from 28-769 mJ/cm² (for Cu). This deviation decreases in thinner films as a result of energy confinement imposed by geometrical constraints. A comparison of the percentage change of LIDT with the thickness between the three noble materials (Figures 6d-8d) and the rest of the materials our results indicate a larger deviation in Au, Ag, Cu. This behavior is attributed to the large ballistic transport lengths of the three materials which cause them to exhibit bulk-like characteristics at larger thicknesses. Another notable effect is the pronounced variation in the maximum lattice temperature $T_L^{max}$ observed at different pulse separations ($t_d = 0, 1, 5, 10$ ps, see Figures 6b–8b), which becomes increasingly significant at larger delays. As shown in the previous section, this trend is absent in materials with lower thermal conductivity and strong electron-phonon coupling. Indeed, the magnitude of these parameters influences the thermal response and gives rise to the observed differences in behavior. More specifically, the delayed equilibration allows an increasing number of electrons to lose energy through enhanced electron diffusion before transferring it to the lattice; this eventually results into lower levels of $T_L^{max}$, significantly smaller than those referred to other materials. As expected, this trend is less intense at smaller $d$ due to energy confinement.

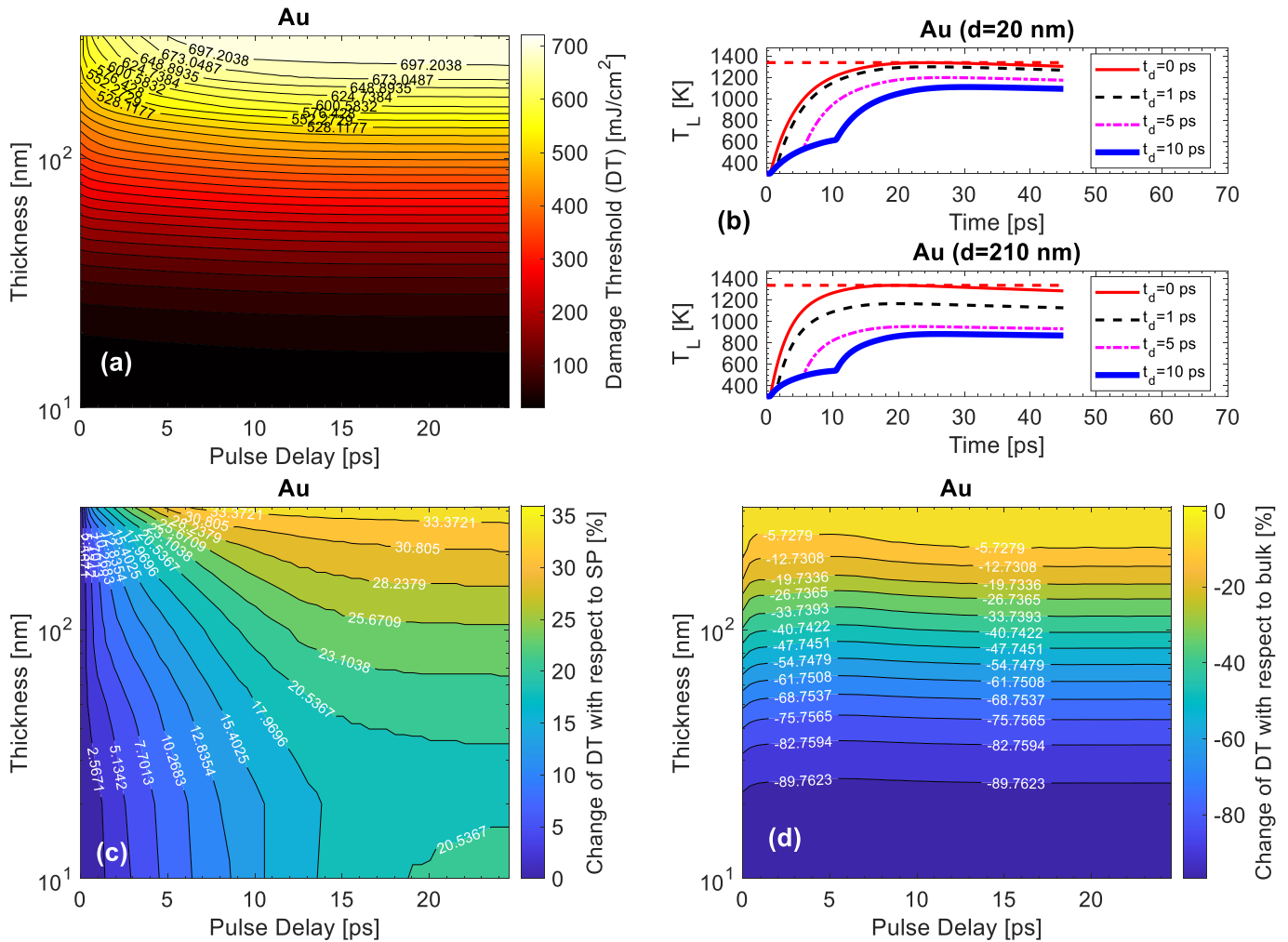

**Figure 6:** Results for Au at different pulse delays and material thicknesses: (a) Damage Threshold (*DT*), (b) Lattice temperature evolution for *d*=20 nm and *d*=210 nm for $F=DT_d^{SP}$ at different time delays $t_d$ (the horizontal *red* dashed line represents the melting point temperature and is shown as a guide to the eye), (c) Percentage change in *DT* relative to single-pulse exposure, and (d) Percentage change in *DT* relative to bulk material response.

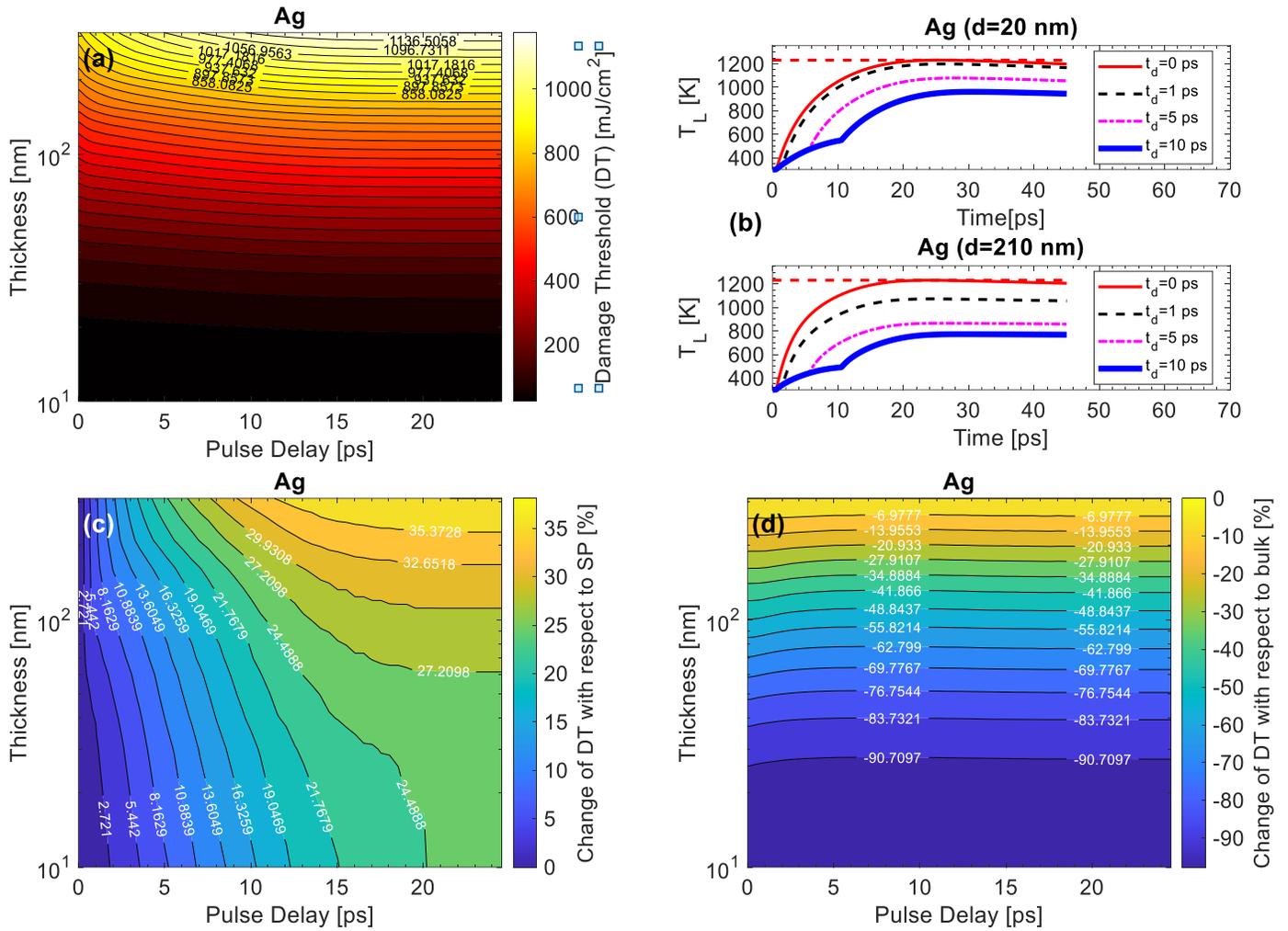

**Figure 7:** Results for Ag at different pulse delays and material thicknesses: (a) Damage Threshold (*DT*), (b) Lattice temperature evolution for *d*=20 nm and *d*=210 nm for $F=DT_d^{SP}$ at different time delays $t_d$ (the horizontal *red* dashed line represents the melting point temperature and is shown as a guide to the eye), (c) Percentage change in *DT* relative to single-pulse exposure, and (d) Percentage change in *DT* relative to bulk material response.

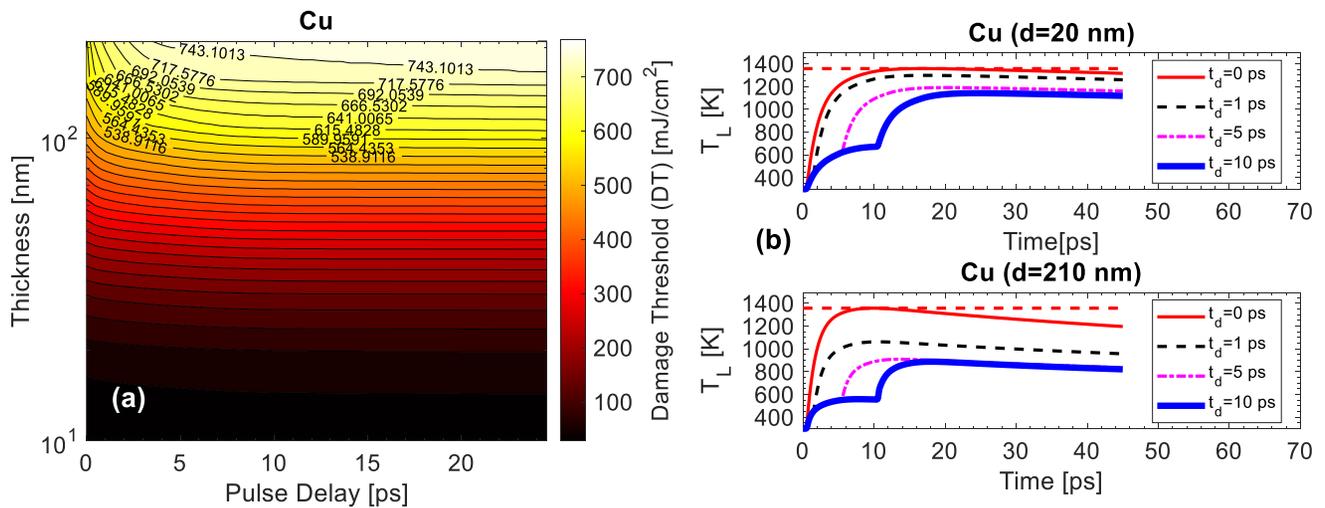

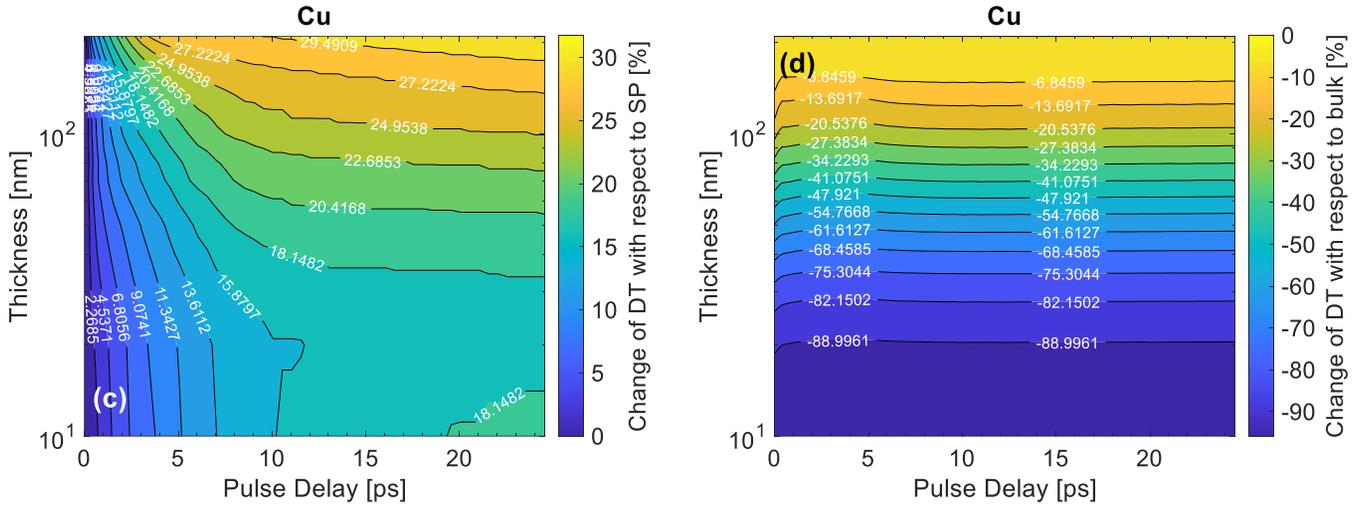

**Figure 8:** Results for Cu at different pulse delays and material thicknesses: (a) Damage Threshold (*DT*), (b) Lattice temperature evolution for *d*=20 nm and *d*=210 nm for $F=DT_d^{SP}$ at different time delays $t_d$ (c) Percentage change in *DT* relative to single-pulse exposure, and (d) Percentage change in *DT* relative to bulk material response.

## 7. Tungsten

Tungsten is a transition metal of moderate electron heat conductivity and high EPC which has a high melting point. Compared to Mo, which has a relatively similar electron-phonon coupling strength but slightly higher electrical conductivity, the calculated LIDT values for the considered range of $t_d$ and $d$ are similar (Figure 9a) and according to the theoretical predictions, LIDT increases with the pulse delay. According to the analysis of the simulated results, W exhibits a slightly greater degree of modulation of LIDT with pulse delay (Figures 9(a,c)) compared to Mo and therefore a greater pulse-delay sensitivity. An interpretation of the noticeable effects in W could be attributed to an expected enhanced electron diffusion affecting, also, the electron-phonon interactions. Results for both thick and thinner (at a smaller degree due to energy confinement) agree with the heat-conductivity-related interpretation. The pronounced role of the higher electron conductivity is reflected in the thermal response of the lattice system at temperature $t_d = 0, 1, 5, 10$ ps (Figure 9b) where the relaxation of the lattice temperatures at all pulse separations drops faster than in W. The decrease of the maximum $T_L^{max}$ with increasing pulse separation reflects the monotonic increase in LIDT. This is confirmed by the analysis of two representative W film thicknesses: $d$=20 nm and $d$=210 nm for which $T_L^{max}$= 3695, 3686, 3668, 3642 K and $T_L^{max}$ = 3695, 3511, 3264, 3131 K (for the four values of $t_d$), respectively. It is noted that for the conditions used in this work, LIDT varies between 25 mJ/cm² and 124 mJ/cm².

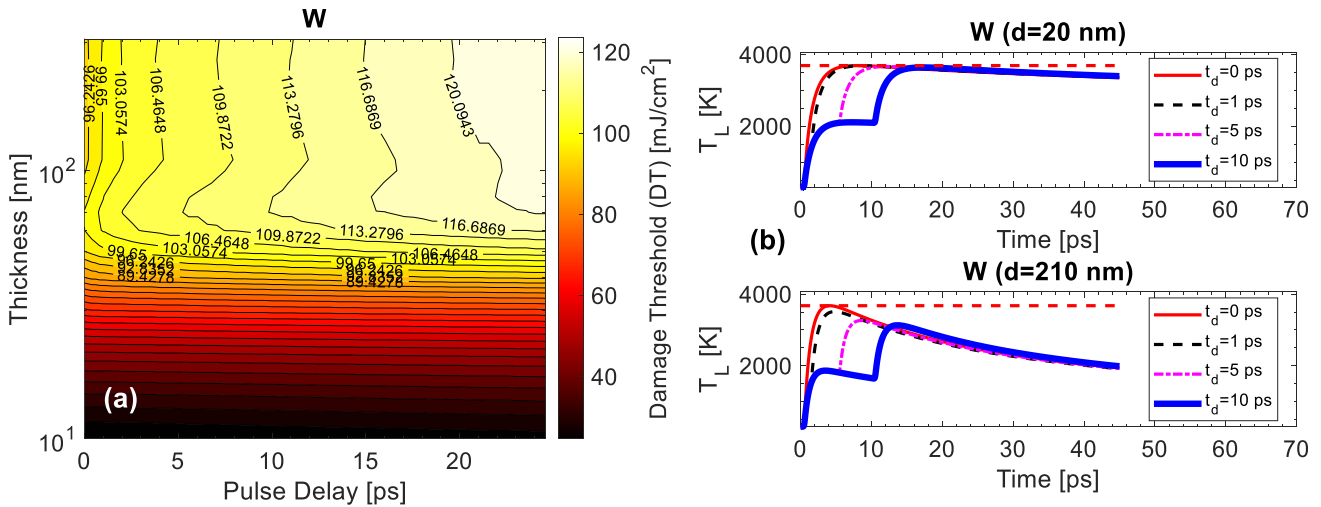

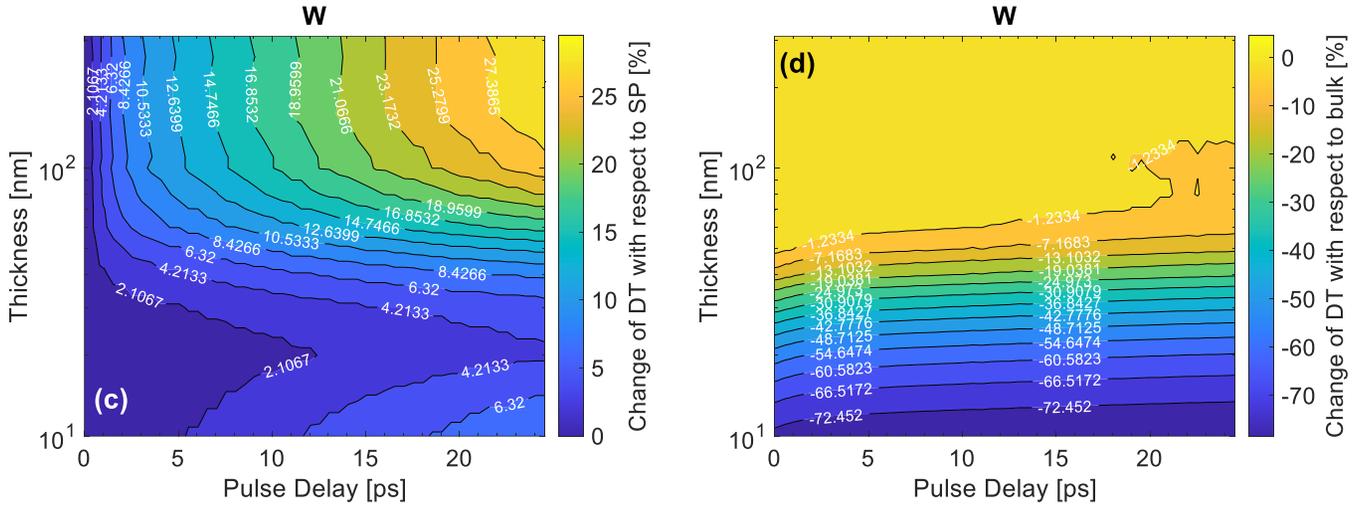

**Figure 9:** Results for W at different pulse delays and material thicknesses: (a) Damage Threshold (*DT*), (b) Lattice temperature evolution for *d*=20 nm and *d*=210 nm for $F=DT_d^{SP}$ at different time delays $t_d$ (the horizontal *red* dashed line represents the melting point temperature and is shown as a guide to the eye), (c) Percentage change in *DT* relative to single-pulse exposure, and (d) Percentage change in *DT* relative to bulk material response.

## 8. Titanium

Titanium is a transition metal with less than half-filled *d*-bands which is characterized with a very low electron heat conductivity and a strong EPC strength. Firstly, the LIDT values for the conditions investigated in this work increases with $t_d$ and it ranges between 11 mJ/cm$^2$ and 58 mJ/cm$^2$ (Figure 10a); this range of values is notably the lowest among all the materials considered in this study. On the other hand, the effect of the remarkably low heat conductivity is reflected on the thermal response of the material and conditions that lead to material melting. According to the results illustrated in Figure 10a, although there is a pronounced dependence of LIDT with the pulse delay, the deviation of LIDT calculated for a $t_d$ from the LIDT for a single pulse does not vary significantly with the thickness (especially, for *d*>20 nm, see Discussion below). This results from the low electron thermal conductivity which has a negligible effect on the dynamical and thermalization behaviour of the material. A second contributing factor is the low excitation level achieved in the material (see Supplementary Material), which, while it is sufficient to induce melting with a single pulse, it is insufficient to facilitate a significant energy transfer to the lattice system.

Thus, the modulation of LIDT with pulse separation should be, rather, attributed to the strength of the electron-phonon coupling, which plays a predominant role in this behavior. An analysis of thermal response of the irradiated material at $t_d$ = 0, 1, 5, 10 ps yield maximum lattice temperatures $T_L^{max}$ = 1941, 1930, 1886, 1849 K (for *d*=210 nm) and $T_L^{max}$ = 1941, 1923, 1885, 1856 K (for *d*=20 nm). These results illustrated in Figure 10b confirm the minimal variation in temperature response (and consequently, the LIDT), which can be attributed to the limited electron conduction.

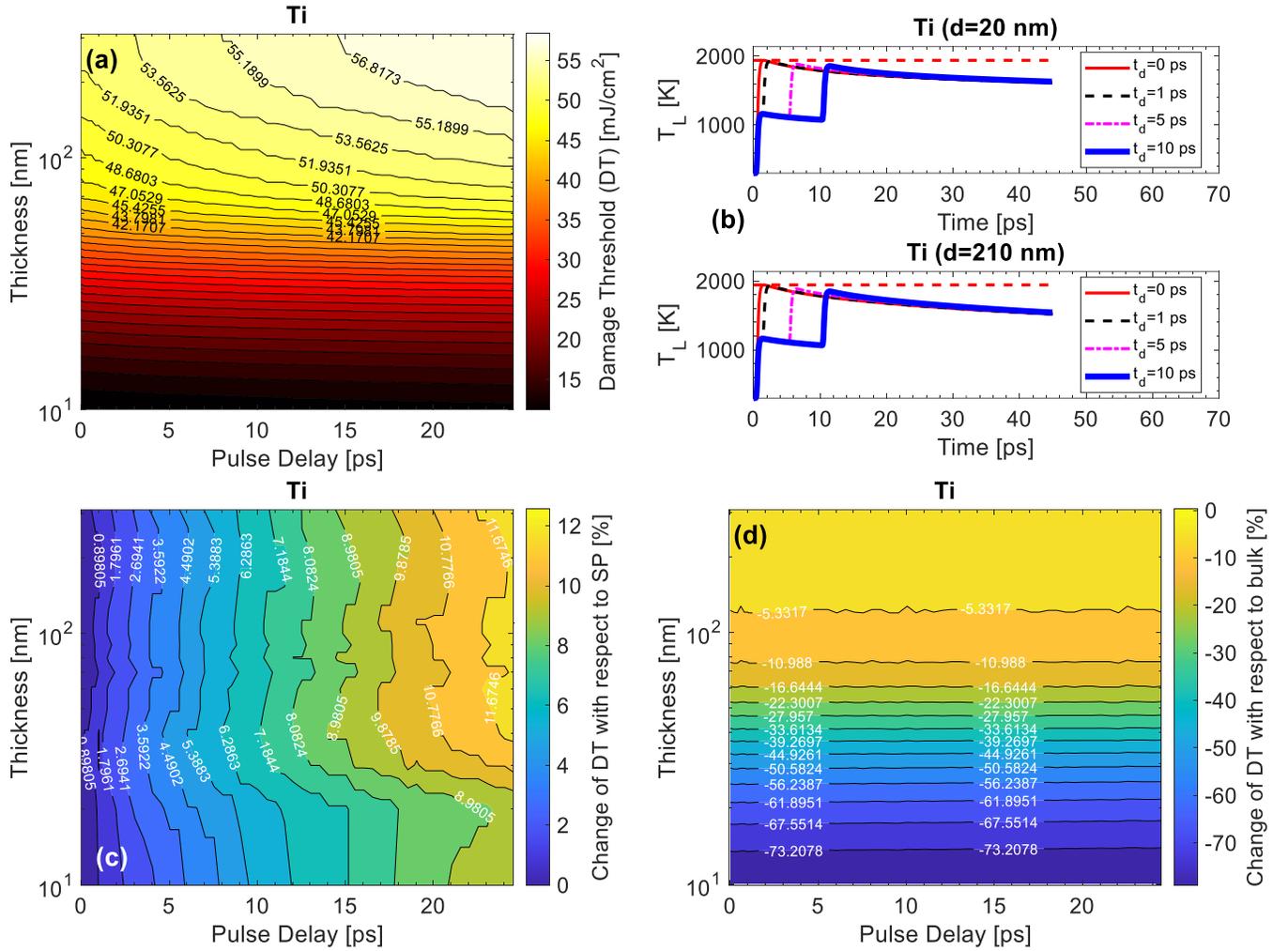

**Figure 10:** Results for Ti at different pulse delays and material thicknesses: (a) Damage Threshold (*DT*), (b) Lattice temperature evolution for *d*=20 nm and *d*=210 nm for $F=DT_d^{SP}$ at different time delays $t_d$ (the horizontal *red* dashed line represents the melting point temperature and is shown as a guide to the eye), (c) Percentage change in *DT* relative to single-pulse exposure, and (d) Percentage change in *DT* relative to bulk material response.

## 9. Steel (100C6)

Steel is an alloy and it is characterized by a relatively low to moderate electron heat conductivity (closer to, but still, a be higher than that of Ti) and a strong EPC strength with an electron temperature-dependence similar to Ni and Pt. As in the case of Ti, the effect of the low heat conductivity is reflected on the thermal response of the material and conditions that lead to material melting. According to the results illustrated in Figure 11a, although there is a pronounced dependence of LIDT with the pulse delay, the deviation of LIDT calculated for a $t_d$ from the LIDT for a single pulse does not vary significantly with the thickness (especially, for *d*>30 nm, see Discussion below). This results from the low electron thermal conductivity which has a negligible effect on the dynamical and thermalization behaviour of the material. Thus, the modulation of LIDT with pulse separation should be, rather, attributed to the strength of the electron-phonon coupling, which plays a predominant role in this behavior. Nevertheless, in contrast to Ti, the relatively higher electron heat conductivity leads to a slightly different profile for the deviation of LIDT calculated for a $t_d$ from the LIDT for a single pulse (Figure 11c) that makes this deviation to be more pronounced at higher *d* than in Ti.

An analysis of thermal response of the irradiated material at $t_d = 0, 1, 5, 10$ ps yield maximum lattice temperatures $T_L^{max} = 1811, 1810, 1763, 1721$ K (for *d*=210 nm) and $T_L^{max} = 1811, 1796, 1760, 1736$ K (for *d*=20 nm). These results confirm the minimal variation in temperature response (and consequently, the LIDT), which can be attributed to the limited electron conduction. At higher pulse separations and smaller thicknesses, the deviation is expected to be larger compared to the comparison with what happens for Ti. In regard to the LIDT values for the conditions investigated in this work, the simulations show that it increases with $t_d$ and it ranges between 18 mJ/cm² and 100 mJ/cm² (Figure 11a).

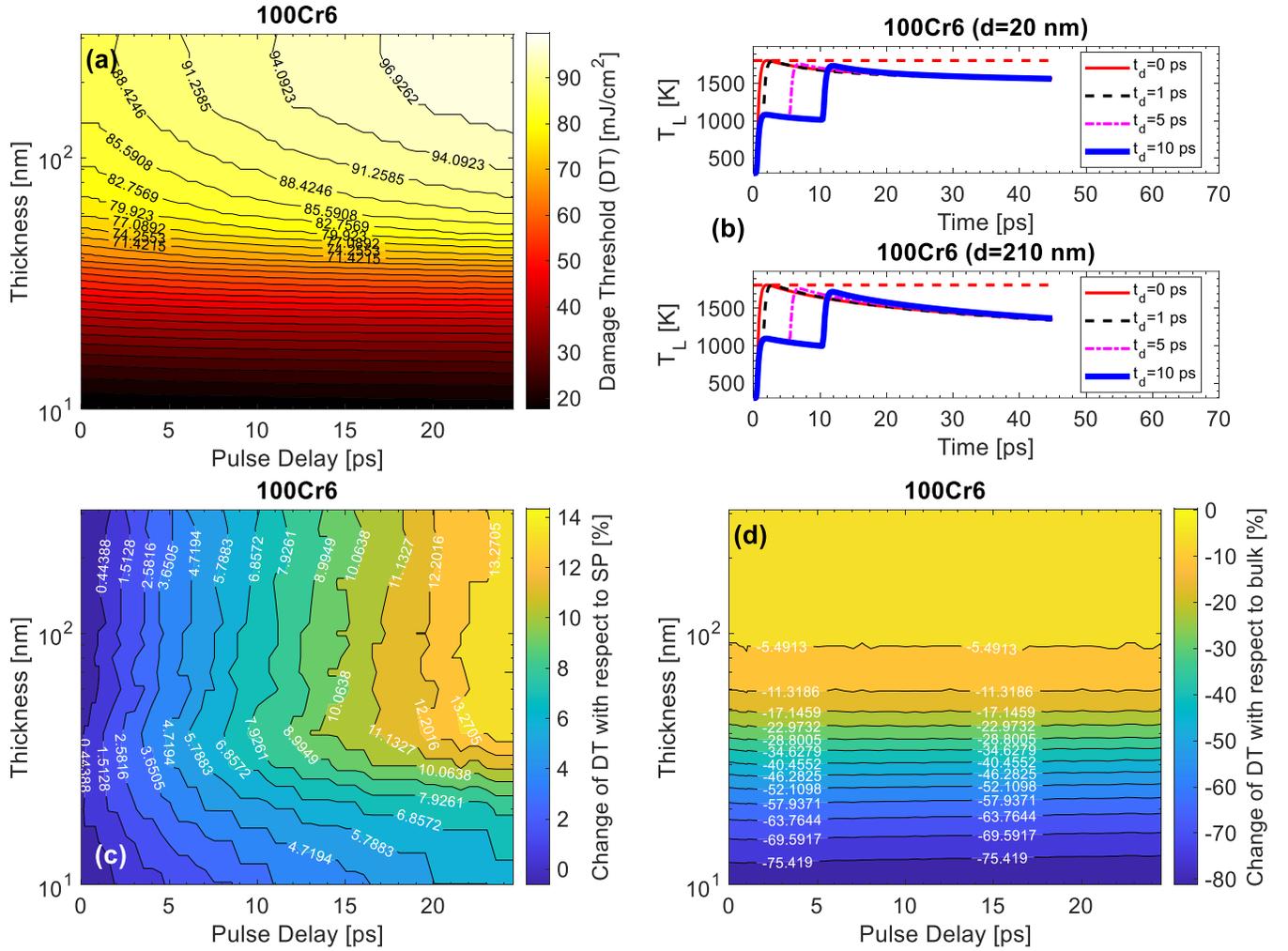

**Figure 11:** Results for 100Cr6 at different pulse delays and material thicknesses: (a) Damage Threshold (*DT*), (b) Lattice temperature evolution for *d*=20 nm and *d*=210 nm for $F=DT_d^{SP}$ at different time delays $t_d$ (the horizontal *red* dashed line represents the melting point temperature and is shown as a guide to the eye), (c) Percentage change in *DT* relative to single-pulse exposure, and (d) Percentage change in *DT* relative to bulk material response.

## B. Optical Properties

It is noted that the LIDT calculations and analysis in the previous sections considered the variations in the optical parameters with both pulse separation and film thickness. In previous reports it was shown both experimentally and theoretically that the optical properties of the irradiated metals vary with film thickness [22]. Results presented in the Supplementary Material demonstrate the evolution of the absorptivity and reflectivity of all materials investigated in this work for two thicknesses, *d*=20 nm and *d*=210 nm and at different pulse separations for $F=DT_d^{SP}$ that corresponds to the laser fluence which causes melting of the material if a single pulse is used for irradiation of a film of thickness *d*. The distinct trend of the optical parameters for the two thicknesses for various materials indicate the impact of interference effects predicted by multireflection theory. In addition to the role of the geometrical constraints, the temporarily deposited and absorbed energy at various inter-pulse delays govern the ultrafast dynamics, and thermal response, including the laser-induced damage threshold (LIDT). The results in the Supplementary Material for reflectivity and absorptivity establish a detailed relationship between the absorbed energy and the excitation level reached in the material, as well as the relaxation time associated with electron-phonon coupling and the electron heat conductivity. For the conditions used in the simulations, the absorptivity values for different inter-pulse delays for the same material thickness, as shown in the Supplementary Material, do not vary significantly; nevertheless, the remarkable LIDT-SP percentage (Figures 1b-11b) changes highlight the crucial role of $t_d$-dependent energy deposition.

## 4. Discussion

The present study provides a systematic theoretical investigation of the laser-induced damage threshold of metallic thin films under double-pulse femtosecond laser irradiation, focusing on the interplay between inter-pulse delay, film thickness, and material-specific thermophysical properties. While the investigation employs the well-established one-dimensional two-temperature model, its novelty arises from the comprehensive, unified approach adopted across eleven technologically relevant metals, including transition metals, noble metals, aluminum, and an industrial alloy. To our knowledge, no prior work has provided such a comparative and self-consistent analysis of LIDT, integrating the ultrafast thermal dynamics, transient optical feedback, and geometrical confinement effects. Furthermore, the study extends previous applications of the 1D-TTM by integrating the thickness-dependent optical properties via multireflection theory, enabling an accurate determination of the transient absorptivity and reflectivity variations. This coupling allows precise calculation of optical feedback effects, which are often neglected in simpler models. Furthermore, the systematic projection of LIDT across varying $t_d$ and $d$ identifies material-specific regimes in which temporal and spatial confinement either synergistically lower the LIDT or partially compensate each other, providing new insights into the mechanisms that dictate laser-induced modification efficiency.

The present investigation showed that across all the investigated materials in this work, short inter-pulse delays result in elevated electron temperatures that transiently modify the optical constants modulating absorptivity. The optical feedback increases the energy retained within the electronic subsystem and, through electron-phonon coupling, intensifies lattice heating. As demonstrated in the analysis, metals with strong electron-phonon coupling exhibit a rapid increase in lattice temperature, whereas those with weaker coupling (or with moderate electron-phonon coupling strength) sustain hot electrons for longer times, extending the sensitivity to the inter-pulse delay $t_d$. Film thickness introduces an additional spatial dimension: when the layer thickness approaches the optical penetration depth, reduced electron diffusion and multiple internal reflections arise from the coupled effects of electronic excitation, electron-phonon coupling, thermal transport, and optical feedback. As demonstrated in Figures 1–11, both the inter-pulse delay $t_d$ and film thickness $d$ characterise the temporal and spatial confinement of energy, thereby defining the conditions that lead to irreversible modification.

Although all studied metals show a pronounced reduction in the LIDT for short $t_d$ and thin films, the extent and persistence of this reduction vary among materials (some metals such as Pt, Ni, or Al exhibit even more complex behaviour for thicker materials), reflecting intrinsic differences in their electronic and thermophysical properties. Overall, the temporal confinement (short $t_d$) and geometrical confinement (small $d$) act synergistically to minimize the energy required for irreversible modification, while high electron conductivity (diffusion) influence these effects. The balance between these competing mechanisms establishes the observed hierarchy of material sensitivity under double femtosecond pulse excitation. These distinctions allow to clearly design rules for tailoring double-pulse femtosecond processing: to enhance modification efficiency, one should employ short $t_d$ and thin films of metals of high electron-phonon coupling strength and low thermal conductivity metals. On the other hand, to minimize damage, longer $t_d$ and thicker films or materials such as W, Mo, Cr, Ti, Pt, Ni, 100Cr6 are preferred. To summarise the discussion on the thermal response of the particular materials, the simulations reveal that LIDT is strongly dictated by the coupled dynamics of electron and lattice subsystems, which are influenced by the electron-phonon coupling strength, electron thermal conductivity, and ballistic transport length of each material. Materials with strong electron-phonon coupling (e.g., Ni, Pt) exhibit rapid lattice heating and a pronounced dependence of LIDT on inter-pulse delay, whereas metals with weak coupling (e.g., Au, Ag, Cu) sustain hot electrons for longer times, extending the sensitivity to $t_d$. The film thickness appears to introduce an additional spatial confinement; more specifically, when $d$ approaches the optical penetration depth, a reduced electron diffusion and multiple internal reflections amplify local heating, which further helpd to modulate LIDT. This marks the synergistic effect of temporal (short $t_d$) and spatial (small $d$) confinement on energy deposition and damage threshold.

The laser-induced damage threshold maps discussed above reveal, also, a remarkable trend (Figures 1c-11c): the deviation of the damage threshold at a specific pulse separation from the single-pulse prediction (LIDT-SP) changes systematically with film thickness (Figures 1c–11c). In particular, it appears that there is a delay in the variation of LIDT-SP at decreasing thickness; the observed delay is attributed to the slower cooling of the lattice at smaller $d$ caused by energy confinement. Interestingly, in some metals (W, Al, Mo, Cr, Pt, Ni) with thicknesses smaller than the optical penetration depth, the behavior does not follow the same trend. From a temperature perspective, this behavior is associated with a more rapid lattice temperature drop; the effect arises predominantly because the energy beam is attenuated within the material, therefore thinner films cannot absorb the entire energy input due to spatial limitations. An analysis of the results illustrated in Figures 1c–11c indicate that materials possessing smaller attenuation profiles (including ballistic length) and lower electron heat conductivity show a stronger manifestation of this behavior.

The above theoretical results reveal the interplay among electronic excitation, energy transfer between the electron and lattice subsystems, and optical feedback in determining the laser-induced damage threshold of metallic thin films. Clearly, an experimental validation is required to confirm the theoretical predictions of the model. In a previous report where the theoretical model was applied to describe ultrafast dynamics following irradiating of thin metal films with *single* pulses ($t_d$=0) [15], results showed a good agreement between theoretical predictions and experimental observations for the damage threshold [15, 18, 42]. Thus, while the present investigation is purely theoretical, it employs a model that has been experimentally validated for single-pulse femtosecond irradiation of thin metal films, which demonstrates an adequate agreement with reported damage thresholds and the ultrafast thermal response [15, 18, 42]. The current work extends this framework to systematically explore the effects of inter-pulse delay and film thickness under double-pulse conditions. Although no new experiments are included here, the predictions provide a detailed theoretical basis for predicting the material response and direct future experimental investigations. Controlled double-pulse femtosecond experiments are therefore essential to confirm these theoretical predictions and to further refine the model for practical applications.

Furthermore, while the simulations in the present work focus on inter-pulse delays up to 25 ps, which correspond to a regime where electron-phonon nonequilibrium and transient optical feedback dominate the material response, it is noted that at longer delays (hundreds of picoseconds to nanoseconds) the system is expected to approach near-equilibrium conditions. In that regime, the electron and lattice subsystems have sufficient time to equilibrate between pulses, and heat diffusion becomes the primary mechanism controlling the temperature evolution. Thus, the lattice temperature and LIDT are expected to gradually converge to the single-pulse limit, effectively decoupling the second pulse from the first. Therefore, the main observations reported in this work are representative of the ultrafast interaction regime, and extending the analysis to the long-delay regime primarily recovers single-pulse behavior rather than introducing new ultrafast dynamics. Nevertheless, the approach can also be extended to describe the thermal response of the system at longer delays.

The substrate is another factor that can influence the thermal response of the system, particularly for thinner films. In a previous study [28], it was demonstrated that opto-thermal behavior differs significantly between thin and thicker films deposited on substrates with different properties, specifically Si and $SiO_2$. Although the substrate thermal conductivity may affect the cooling dynamics and variations in the substrate refractive index can modify the film absorptivity, a systematic investigation of their impact on the laser-induced damage threshold is beyond the scope of the present work. Nonetheless, the theoretical framework employed in the current study can still employed for other substrates.

In conlusion, a significant contribution of this work is the establishment of correlations between fundamental material parameters such as the electron-phonon coupling strength, electron thermal conductivity, optical parameters, ballistic transport length and the observed LIDT behaviour. These correlations form the basis for design rules in double-pulse femtosecond processing: these results manifested that, high electron-phonon coupling strengths and low thermal conductivity metals are more sensitive to short $t_d$ and thin films, enabling efficient material modification, while metals with higher conductivity or thicker films are more robust, minimizing an undesired damage. In addition, by presenting a comparative LIDT database for eleven metals, these results provide a comprehensive reference for optimizing double-pulse femtosecond laser interactions in metallic thin films, with implications for laser micromachining, surface engineering, and materials processing. The database provides not only the absolute thresholds but also the percentage deviations relative to single-pulse irradiation, offering guidance on tailoring pulse sequences and film thickness to achieve specific material modifications. Overall, this work goes beyond a simple application of the TTM by providing a comprehensive, cross-material, and thickness-resolved analysis that links ultrafast electron-lattice dynamics with practical material response under double-pulse irradiation. The study clarifies the competing roles of temporal and spatial confinement, identifies material-dependent sensitivities, and establishes a framework that can guide both fundamental studies and industrial applications. These novel insights emphasises the value of a unified theoretical treatment in understanding and optimizing laser-metal interactions for femtosecond processing.

## 5. Conclusions

In this work, we systematically investigated the influence of inter-pulse delay and film thickness on the laser-induced damage threshold of metallic thin films under double femtosecond pulse irradiation. The results demonstrate that both temporal and geometrical confinement strongly modulate energy deposition and material response. Short inter-pulse delays elevate electron temperatures, transiently modify optical properties, and enhance lattice heating through electron-phonon coupling, reducing LIDT. Thin films exhibit pronounced thermal confinement, further lowering the damage threshold. Material-specific trends are characterised by the interplay of electron thermal conductivity, electron-phonon coupling strength, and film thickness, dictating sensitivity to pulse separation. These insights provide particular guidelines for tailoring double-pulse femtosecond laser processing: short delays and thin films maximize modification efficiency, whereas longer delays, thicker films, or materials with high thermal conductivity and moderate electron-phonon coupling minimize damage. Overall, the study establishes a predictive framework for designing ultrafast laser processing strategies across a wide range of metallic systems and set the theoretical basis for experimental validation and optimized micro- and nano-fabrication protocols.

**Declaration of competing interest**

The authors declare that they have no known competing financial interests or personal relationships that could have appeared to influence the work reported in this paper.

**Acknowledgements**

This research work was supported by the EU-Horizon 2020 Nanoscience Foundries and Fine Analysis (NEP) Project (Grant agreement ID: 101007417) and from Horizon Europe, the European Union's Framework Programme for Research and Innovation, under Grant Agreement No. 101057457 (METAMORPHA) and European Projects Lasers4EU (Grant agreement ID: 101131771) and RIANA (Grant agreement ID: 101130652).

**Appendix A. Supplementary data**

Supplementary data to this article can be found online at https://doi. org/.

**Data availability**

The datasets generated and analyzed during the current study are described in full within the manuscript and Supplementary Material. On the other hand, the numerical code used for simulations is not publicly released. Requests for access to the code may include guidance on extending the model to modify parameters such as other material addition, substrate thermal properties, laser wavelength, pulse duration, or film thickness. This approach ensures reproducibility while allowing controlled use and collaboration. Earlier versions of the computational approach can be found in https://nffa.eu/apply/virtual-access-services/modelling-and-machine-learning/ while an interface enabling users to explore optothermal effects across different laser conditions, materials, and substrates will be released in the coming months as part of the European projects RIANA and Lasers4EU.

**References**

[1] E. Stratakis, J. Bonse, J. Heitz, J. Siegel, G.D. Tsibidis, E. Skoulas, A. Papadopoulos, A. Mimidis, A.C. Joel, P. Comanns, J. Krüger, C. Florian, Y. Fuentes-Edfuf, J. Solis, W. Baumgartner, Laser engineering of biomimetic surfaces, Materials Science and Engineering: R: Reports 141 (2020) 100562.
[2] F. Fraggelakis, P. Lingos, G.D. Tsibidis, E. Cusworth, N. Kay, L. Fumagalli, V.G. Kravets, A.N. Grigorenko, A.V. Kabashin, E. Stratakis, Double-Pulse Femtosecond Laser Fabrication of Highly Ordered Periodic Structures on Au Thin Films Enabling Low-Cost Plasmonic Applications, ACS Nano 19(25) (2025) 23258-23275.
[3] G.D. Tsibidis, L. Museur, A. Kanaev, The Role of Crystalline Orientation in the Formation of Surface Patterns on Solids Irradiated with Femtosecond Laser Double Pulses, Applied Sciences 10(24) (2020).
[4] F. Fraggelakis, G. Mincuzzi, J. Lopez, I. Manek-Honninger, R. Kling, Controlling 2D laser nano structuring over large area with double femtosecond pulses, Applied Surface Science 470 (2019) 677-686.
[5] F. Fraggelakis, G. Giannuzzi, C. Gaudiuso, I. Manek-Honninger, G. Mincuzzi, A. Ancona, R. Kling, Double- and Multi-Femtosecond Pulses Produced by Birefringent Crystals for the Generation of 2D Laser-Induced Structures on a Stainless Steel Surface, Materials 12(8) (2019).

[6] S. Hohm, A. Rosenfeld, J. Kruger, J. Bonse, Laser-induced periodic surface structures on titanium upon single- and two-color femtosecond double-pulse irradiation, Optics Express 23(20) (2015) 25959-25971.
[7] S. Hohm, M. Herzlieb, A. Rosenfeld, J. Kruger, J. Bonse, Femtosecond laser-induced periodic surface structures on silicon upon polarization controlled two-color double-pulse irradiation, Optics Express 23(1) (2015) 61-71.
[8] G.D. Tsibidis, Thermal response of double-layered metal films after ultrashort pulsed laser irradiation: The role of nonthermal electron dynamics, Applied Physics Letters 104(5) (2014) 051603.
[9] T.J.Y. Derrien, J. Kruger, T.E. Itina, S. Hohm, A. Rosenfeld, J. Bonse, Rippled area formed by surface plasmon polaritons upon femtosecond laser double-pulse irradiation of silicon: the role of carrier generation and relaxation processes, Applied Physics a-Materials Science & Processing 117(1) (2014) 77-81.
[10] S. Hohm, A. Rosenfeld, J. Kruger, J. Bonse, Area dependence of femtosecond laser-induced periodic surface structures for varying band gap materials after double pulse excitation, Applied Surface Science 278 (2013) 7-12.
[11] T.J.Y. Derrien, J. Kruger, T.E. Itina, S. Höhm, A. Rosenfeld, J. Bonse, Rippled area formed by surface plasmon polaritons upon femtosecond laser double-pulse irradiation of silicon, Optics Express 21(24) (2013) 29643-29655.
[12] M. Barberoglou, G.D. Tsibidis, D. Gray, E. Magoulakis, C. Fotakis, E. Stratakis, P.A. Loukakos, The influence of ultra-fast temporal energy regulation on the morphology of Si surfaces through femtosecond double pulse laser irradiation, Applied Physics A: Materials Science and Processing 113 (2013) 273-283.
[13] M. Barberoglou, D. Gray, E. Magoulakis, C. Fotakis, P.A. Loukakos, E. Stratakis, Controlling ripples' periodicity using temporally delayed femtosecond laser double pulses, Optics Express 21(15) (2013) 18501-18508.
[14] I.H. Chowdhury, X.F. Xu, A.M. Weiner, Ultrafast double-pulse ablation of fused silica, Applied Physics Letters 86(15) (2005).
[15] G.D. Tsibidis, D. Mansour, E. Stratakis, Damage threshold evaluation of thin metallic films exposed to femtosecond laser pulses: The role of material thickness, Opt Laser Technol 156 (2022) 108484.
[16] J. Hohlfeld, S.S. Wellershoff, J. Güdde, U. Conrad, V. Jahnke, E. Matthias, Electron and lattice dynamics following optical excitation of metals, Chemical Physics 251(1-3) (2000) 237-258.
[17] M. Bonn, D.N. Denzler, S. Funk, M. Wolf, S.S. Wellershoff, J. Hohlfeld, Ultrafast electron dynamics at metal surfaces: Competition between electron-phonon coupling and hot-electron transport, Physical Review B 61(2) (2000) 1101-1105.
[18] S.S. Wellershoff, J. Hohlfeld, J. Güdde, E. Matthias, The role of electron–phonon coupling in femtosecond laser damage of metals, Applied Physics A 69(1) (1999) S99-S107.
[19] J. Güdde, J. Hohlfeld, J.G. Müller, E. Matthias, Damage threshold dependence on electron–phonon coupling in Au and Ni films, Applied Surface Science 127 (1998) 40-45.
[20] J. Hohlfeld, D. Grosenick, U. Conrad, E. Matthias, Femtosecond Time-Resolved Reflection 2nd-Harmonic Generation on Polycrystalline Copper, Applied Physics a-Materials Science & Processing 60(2) (1995) 137-142.
[21] M.-C. Velli, S. Maragkaki, M. Vlahou, G.D. Tsibidis, E. Stratakis, Investigation of the role of pulse duration and film thickness on the damage threshold of metal thin films irradiated with ultrashort laser pulses, Applied Surface Science 657 (2024) 159810.
[22] J. Hohlfeld, J.G. Muller, S.S. Wellershoff, E. Matthias, Time-resolved thermoreflectivity of thin gold films and its dependence on film thickness, Applied Physics B-Lasers and Optics 64(3) (1997) 387-390.
[23] B.Y. Mueller, B. Rethfeld, Nonequilibrium electronβ€"phonon coupling after ultrashort laser excitation of gold, Applied Surface Science (http://dx.doi.org/10.1016/j.apsusc.2013.12.074 ) (0) (2014).
[24] D.S. Ivanov, B. Rethfeld, The effect of pulse duration on the interplay of electron heat conduction and electron-phonon interaction: Photo-mechanical versus photo-thermal damage of metal targets, Applied Surface Science 255(24) (2009) 9724-9728.
[25] J.P. Colombier, F. Garrelie, N. Faure, S. Reynaud, M. Bounhalli, E. Audouard, R. Stoian, F. Pigeon, Effects of electron-phonon coupling and electron diffusion on ripples growth on ultrafast-laser-irradiated metals, Journal of Applied Physics 111(2) (2012) 024902.
[26] B. Rethfeld, M.E. Garcia, D.S. Ivanov, S. Anisimov, Modelling ultrafast laser ablation, Journal of Physics D: Applied Physics 50(19) (2017) 193001.
[27] S.I. Anisimov, B.L. Kapeliovich, T.L. Perel'man, Electron-emission from surface of metals induced by ultrashort laser pulses, Zhurnal Eksperimentalnoi Teor. Fiz. 66(2) (1974 [Sov. Phys. Tech. Phys. 11, 945 (1967)]) 776-781.
[28] G.D. Tsibidis, E. Stratakis, Impact of substrate on opto-thermal response of thin metallic targets under irradiation with femtosecond laser pulses, J Cent South Univ 29(10) (2022) 3410-3421.
[29] A.M. Chen, L.Z. Sui, Y. Shi, Y.F. Jiang, D.P. Yang, H. Liu, M.X. Jin, D.J. Ding, Ultrafast investigation of electron dynamics in the gold-coated two-layer metal films, Thin Solid Films 529 (2013) 209-216.
[30] A.M. Chen, H.F. Xu, Y.F. Jiang, L.Z. Sui, D.J. Ding, H. Liu, M.X. Jin, Modeling of femtosecond laser damage threshold on the two-layer metal films, Applied Surface Science 257(5) (2010) 1678-1683.
[31] M. Born, E. Wolf, Principles of optics : electromagnetic theory of propagation, interference and diffraction of light, 7th expanded ed., Cambridge University Press, Cambridge ; New York, 1999.

[32] I.H. Malitson, Interspecimen Comparison of Refractive Index of Fused Silica, Journal of the Optical Society of America 55(10P1) (1965) 1205-1209.
[33] A.D. Rakic, A.B. Djurisic, J.M. Elazar, M.L. Majewski, Optical Properties of Metallic Films for Vertical-Cavity Optoelectronic Devices, Applied Optics 37(22) (1998) 5271-5283.
[34] G.D. Tsibidis, A. Mimidis, E. Skoulas, S.V. Kirner, J. Krüger, J. Bonse, E. Stratakis, Modelling periodic structure formation on 100Cr6 steel after irradiation with femtosecond-pulsed laser beams, Applied Physics A 124(1) (2017) 27.
[35] M.A. Ordal, R.J. Bell, R.W. Alexander, L.A. Newquist, M.R. Querry, Optical properties of Al, Fe, Ti, Ta, W, and Mo at submillimeter wavelengths, Applied Optics 27(6) (1988) 1203-1209.
[36] Z. Lin, L.V. Zhigilei, V. Celli, Electron-phonon coupling and electron heat capacity of metals under conditions of strong electron-phonon nonequilibrium, Physical Review B 77(7) (2008) 075133.
[37] F. Fraggelakis, G.D. Tsibidis, E. Stratakis, Tailoring submicrometer periodic surface structures via ultrashort pulsed direct laser interference patterning, Physical Review B 103(5) (2021) 054105.
[38] B.H. Christensen, K. Vestentoft, P. Balling, Short-pulse ablation rates and the two-temperature model, Applied Surface Science 253(15) (2007) 6347-6352.
[39] S. Petrovic, G.D. Tsibidis, A. Kovacevic, N. Bozinovic, D. Perusko, A. Mimidis, A. Manousaki, E. Stratakis, Effects of static and dynamic femtosecond laser modifications of Ti/Zr multilayer thin films, Eur Phys J D 75(12) (2021) 304.
[40] D.R. Lide, CRC Handbook of Chemistry and Physics, 84th Edition, 2003-2004).
[41] L.V. Zhigilei, Database on 'Electron-Phonon Coupling and Electron Heat Capacity in Metals at High Electron Temperatures' (https://faculty.virginia.edu/CompMat/electron-phonon-coupling/), 2003.
[42] B.C. Stuart, M.D. Feit, S. Herman, A.M. Rubenchik, B.W. Shore, M.D. Perry, Optical ablation by high-power short-pulse lasers, J. Opt. Soc. Am. B 13(2) (1996) 459-468.

# Supplementary Material to

*'Influence of Inter-Pulse Delay and Geometric Constraints on Damage and Optical Characteristics in thin Metal Targets Irradiated by Double Ultrashort Laser Pulses'*

George D. Tsibidis[1,a]

[1]*Institute of Electronic Structure and Laser (IESL), Foundation for Research and Technology (FORTH), Vassilika Vouton, 70013, Heraklion, Crete, Greece*

[a] Authors to whom correspondence should be addressed: tsibidis@iesl.forth.gr

## 1. Opto-thermal response of irradiated solids

In Figures SM1-11, the evolution of electron and lattice temperatures for eleven metals (Au, Ag, Cu, Al, Ni, Ti, Cr, Pt, W, Mo, and Stainless Steel (100Cr6)) of thicknesses $d$=20 nm (a) and $d$=210 nm (b) are calculated at fluences $F=DT_d^{SP}$ for various pulse separations. $DT_d^{SP}$ denotes the damage threshold under single-pulse irradiation for a metal film of thickness $d$. Similarly, for the same pulse separations and material thicknesses, the evolution of the reflectivity (c,d) and absorptivity (e,f) are illustrated. It is noted, though, the optical property evolution is calculated for all metals except from Mo as, to the best of our knowledge, there is no previous work that reports a precise expression for the calculation of the dielectric parameter of Mo based on the Drude-Lorentz model.

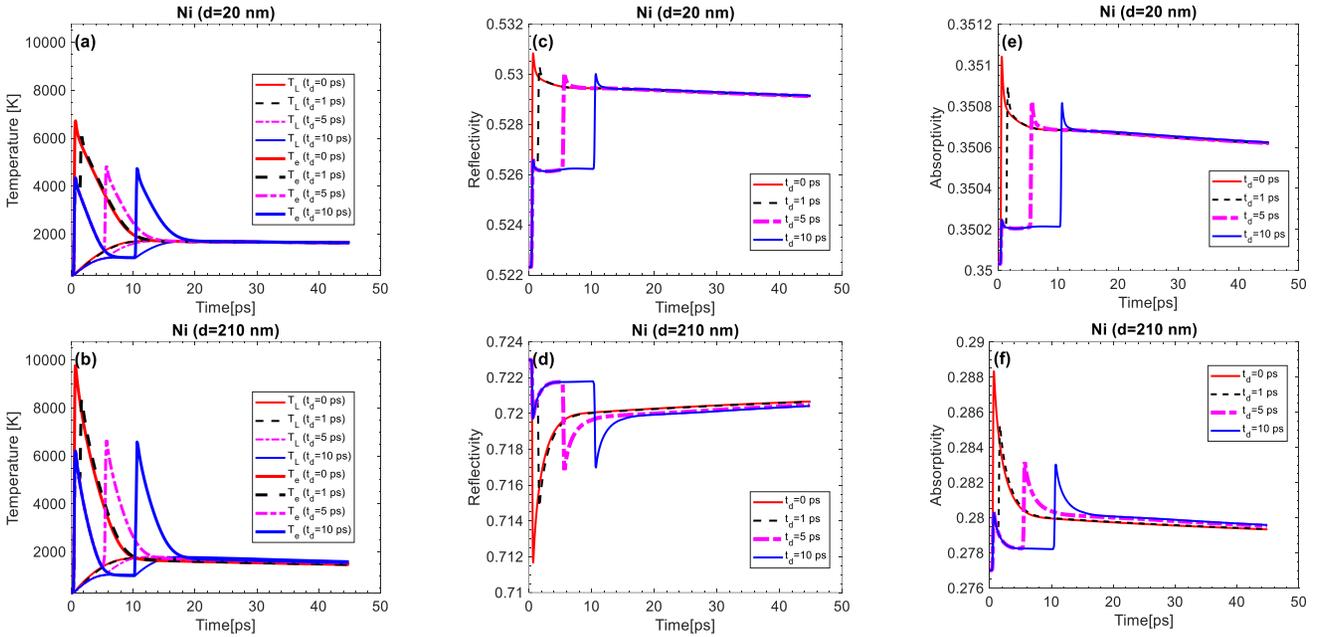

**Figure SM1:** Results for Ni at different pulse delays and material thicknesses: Electron and Lattice Temperature evolution at $d$=20 nm (a) and $d$=210 nm (b) for $F=DT_d^{SP}$ at different time delays $t_d$; Reflectivity evolution at $d$=20 nm (c) and $d$=210 nm (d) for $F=DT_d^{SP}$ at different time delays $t_d$ (c); Absorptivity evolution at $d$=20 nm (e) and $d$=210 nm (f) for $F=DT_d^{SP}$ at different time delays $t_d$.

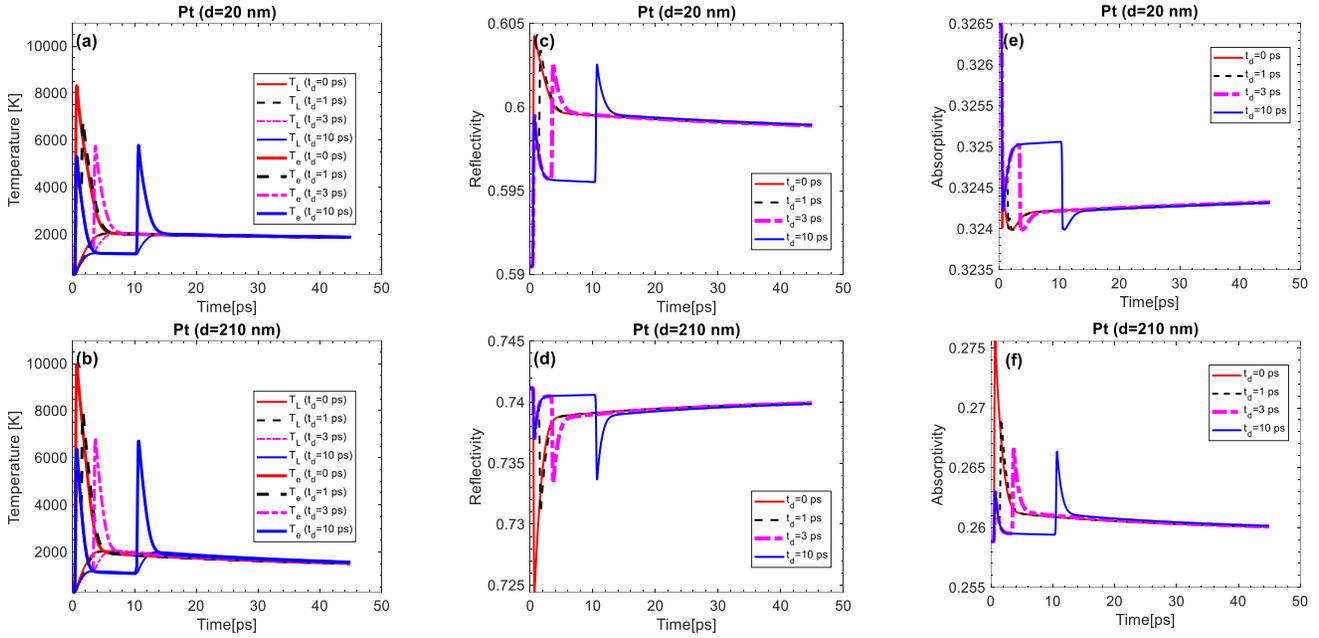

**Figure SM2:** Results for Pt at different pulse delays and material thicknesses: Electron and Lattice Temperature evolution at $d$=20 nm (a) and $d$=210 nm (b) for $F=DT_d^{SP}$ at different time delays $t_d$; Reflectivity evolution at $d$=20 nm (c) and $d$=210 nm (d) for $F=DT_d^{SP}$ at different time delays $t_d$ (c); Absorptivity evolution at $d$=20 nm (e) and $d$=210 nm (f) for $F=DT_d^{SP}$ at different time delays $t_d$.

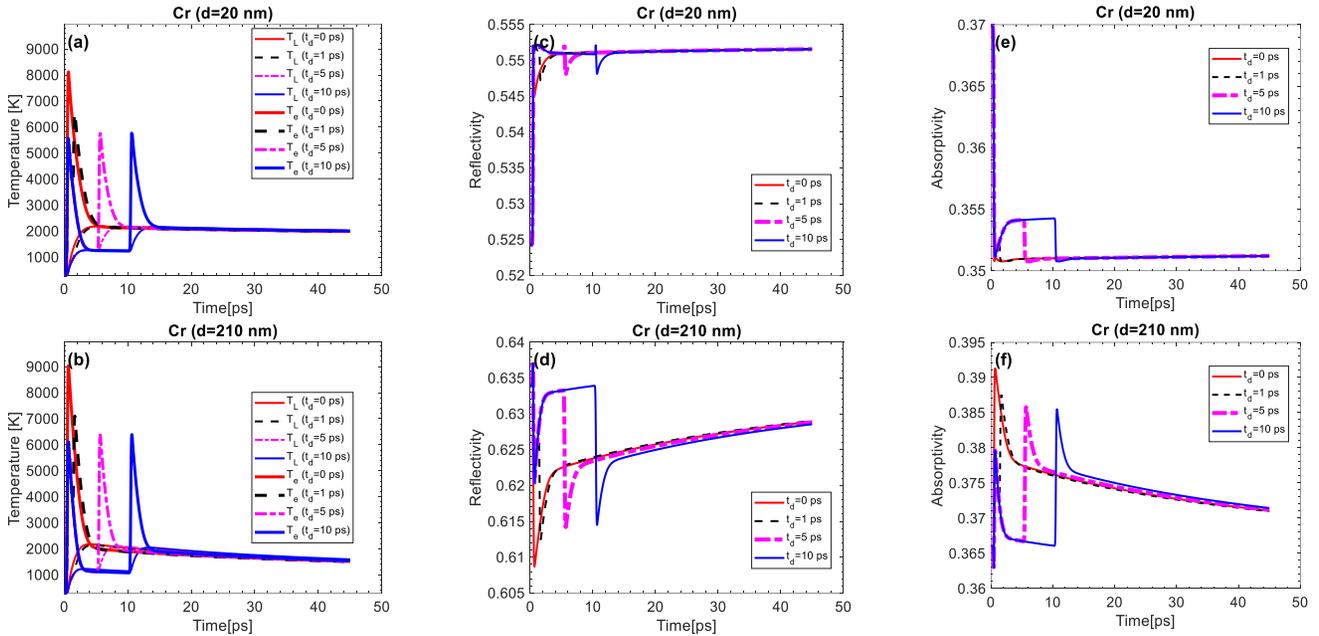

**Figure SM3:** Results for Cr at different pulse delays and material thicknesses: Electron and Lattice Temperature evolution at $d$=20 nm (a) and $d$=210 nm (b) for $F=DT_d^{SP}$ at different time delays $t_d$; Reflectivity evolution at $d$=20 nm (c) and $d$=210 nm (d) for $F=DT_d^{SP}$ at different time delays $t_d$ (c); Absorptivity evolution at $d$=20 nm (e) and $d$=210 nm (f) for $F=DT_d^{SP}$ at different time delays $t_d$.

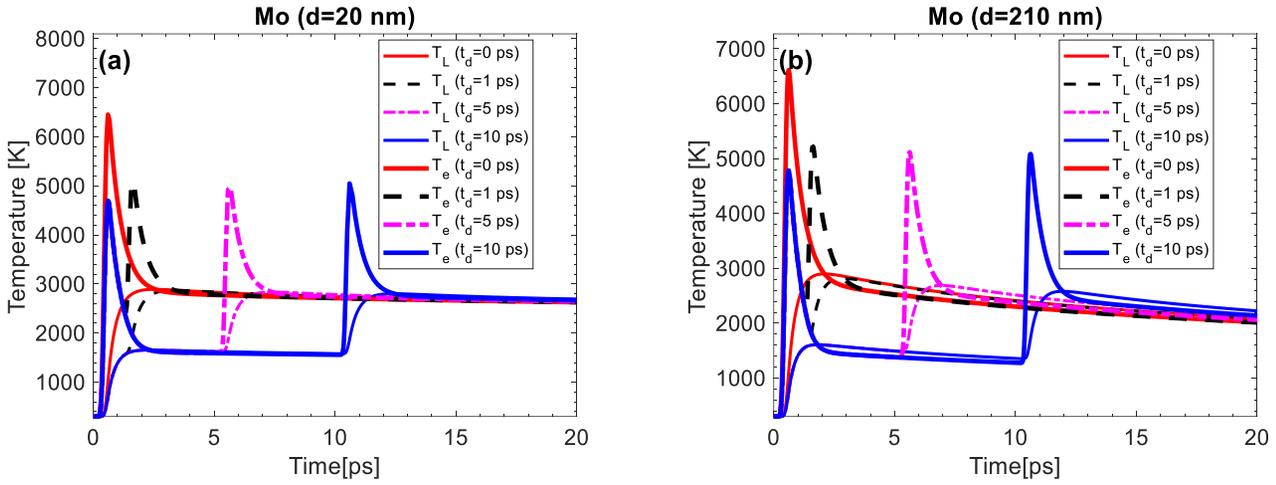

**Figure SM4:** Results for Mo at different pulse delays and material thicknesses: Electron and Lattice Temperature evolution at $d$=20 nm (a) and $d$=210 nm (b) for $F=DT_d^{SP}$ at different time delays $t_d$. Reflectivity is calculated to be 0.4158 (for $d$=20 nm) and 0.666 (for $d$=210 nm) while absorptivity is calculated to be 0.3796 (for $d$=20 nm) and 0.334 (for $d$=210 nm).

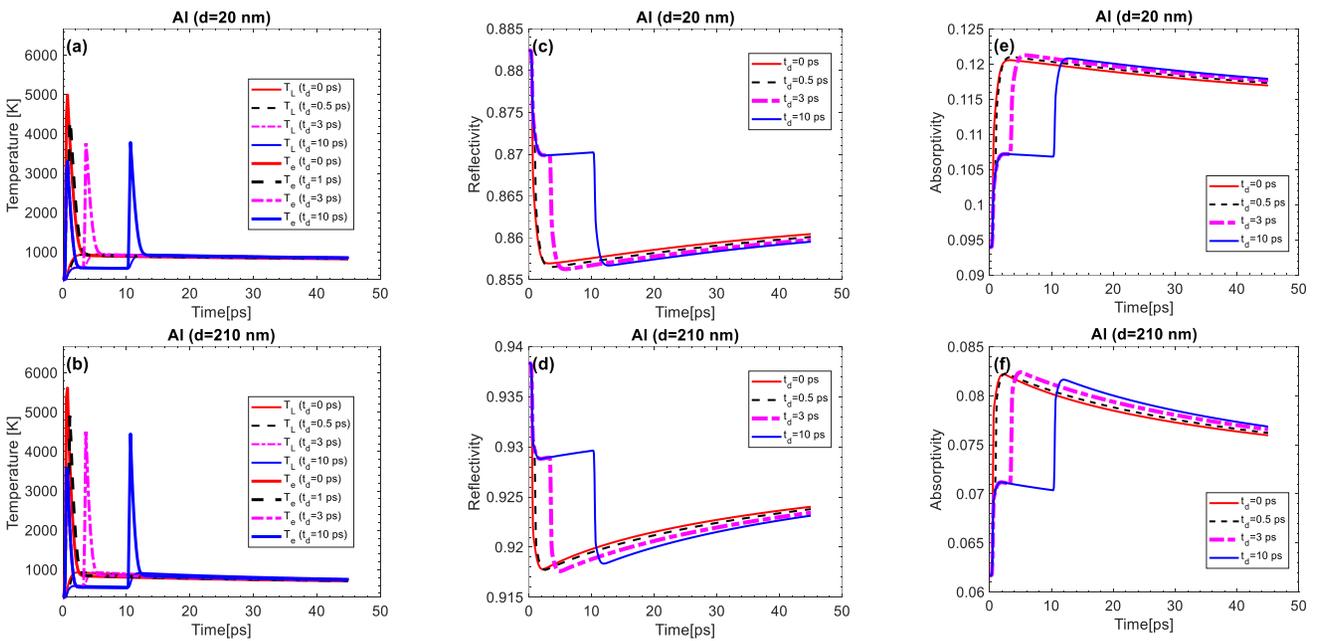

**Figure SM5:** Results for Al at different pulse delays and material thicknesses: Electron and Lattice Temperature evolution at $d$=20 nm (a) and $d$=210 nm (b) for $F=DT_d^{SP}$ at different time delays $t_d$; Reflectivity evolution at $d$=20 nm (c) and $d$=210 nm (d) for $F=DT_d^{SP}$ at different time delays $t_d$ (c); Absorptivity evolution at $d$=20 nm (e) and $d$=210 nm (f) for $F=DT_d^{SP}$ at different time delays $t_d$.

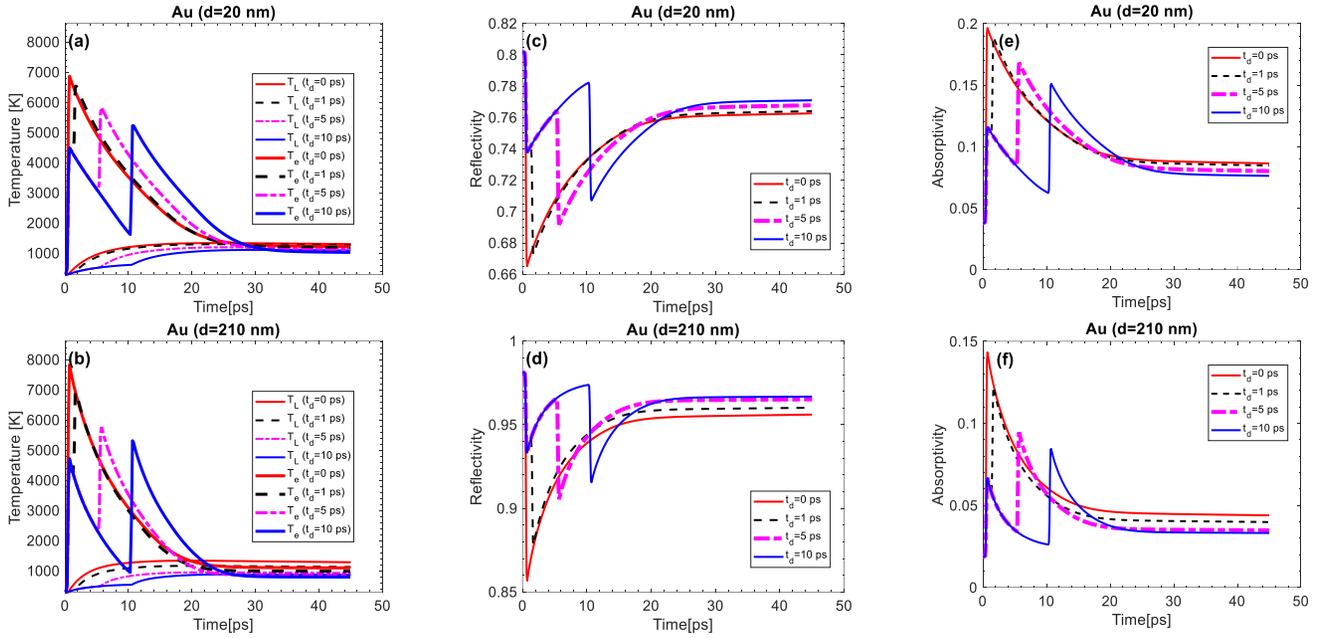

**Figure SM6:** Results for Au at different pulse delays and material thicknesses: Electron and Lattice Temperature evolution at $d=20$ nm (a) and $d=210$ nm (b) for $F=DT_d^{SP}$ at different time delays $t_d$; Reflectivity evolution at $d=20$ nm (c) and $d=210$ nm (d) for $F=DT_d^{SP}$ at different time delays $t_d$ (c); Absorptivity evolution at $d=20$ nm (e) and $d=210$ nm (f) for $F=DT_d^{SP}$ at different time delays $t_d$.

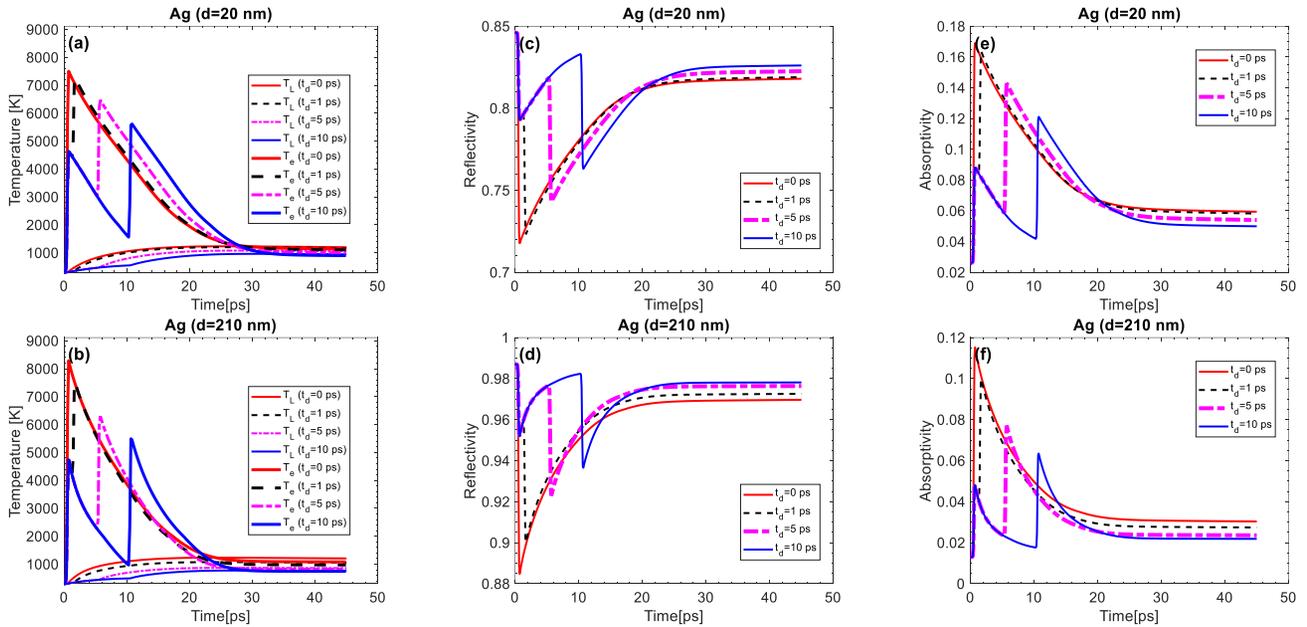

**Figure SM7:** Results for Ag at different pulse delays and material thicknesses: Electron and Lattice Temperature evolution at $d=20$ nm (a) and $d=210$ nm (b) for $F=DT_d^{SP}$ at different time delays $t_d$; Reflectivity evolution at $d=20$ nm (c) and $d=210$ nm (d) for $F=DT_d^{SP}$ at different time delays $t_d$ (c); Absorptivity evolution at $d=20$ nm (e) and $d=210$ nm (f) for $F=DT_d^{SP}$ at different time delays $t_d$.

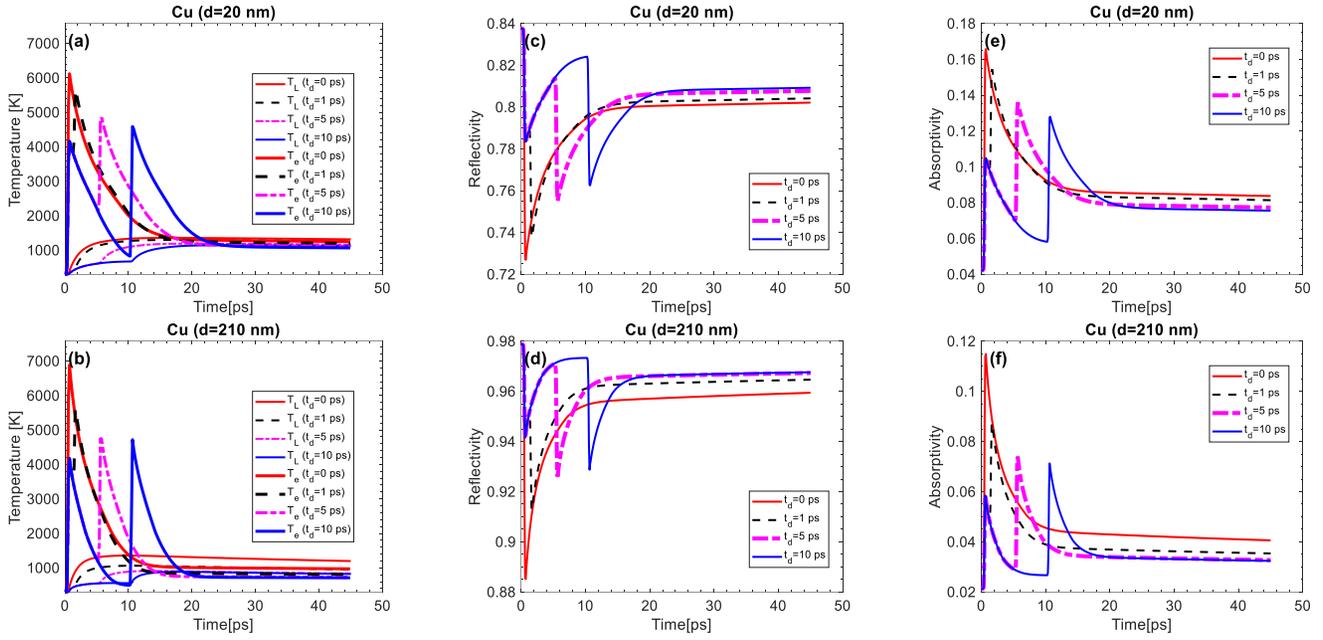

**Figure SM8:** Results for Cu at different pulse delays and material thicknesses: Electron and Lattice Temperature evolution at $d$=20 nm (a) and $d$=210 nm (b) for $F=DT_d^{SP}$ at different time delays $t_d$; Reflectivity evolution at $d$=20 nm (c) and $d$=210 nm (d) for $F=DT_d^{SP}$ at different time delays $t_d$ (c); Absorptivity evolution at $d$=20 nm (e) and $d$=210 nm (f) for $F=DT_d^{SP}$ at different time delays $t_d$.

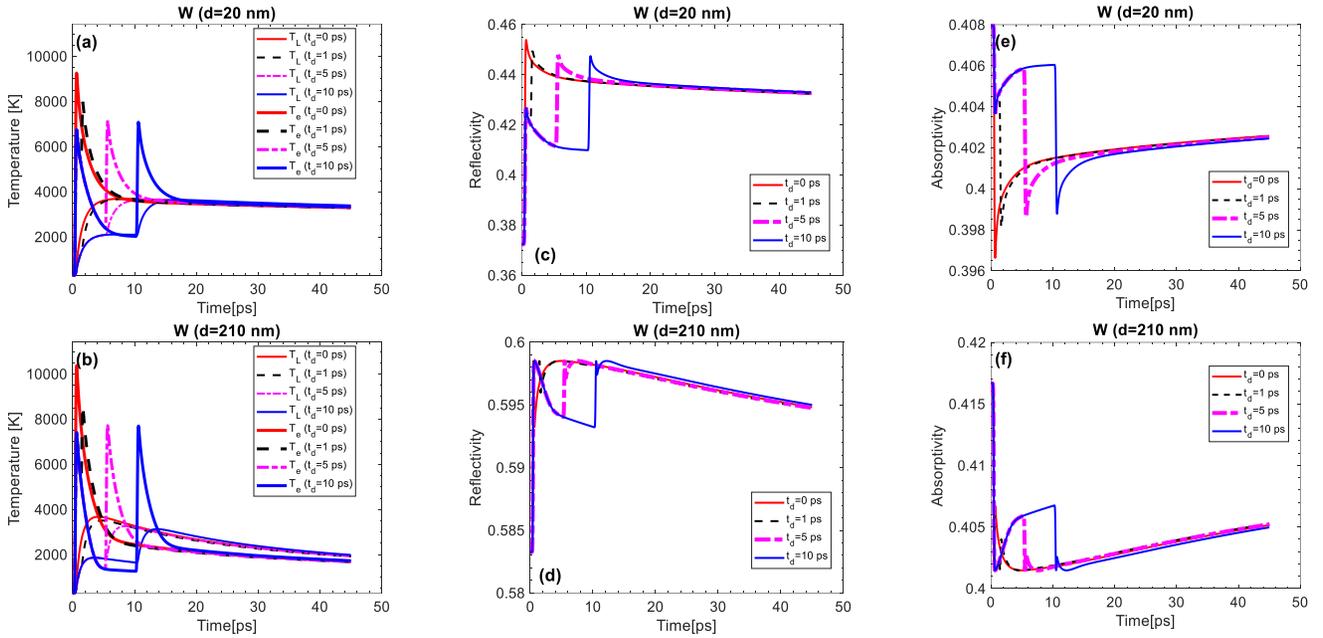

**Figure SM9:** Results for W at different pulse delays and material thicknesses: Electron and Lattice Temperature evolution at $d$=20 nm (a) and $d$=210 nm (b) for $F=DT_d^{SP}$ at different time delays $t_d$; Reflectivity evolution at $d$=20 nm (c) and $d$=210 nm (d) for $F=DT_d^{SP}$ at different time delays $t_d$ (c); Absorptivity evolution at $d$=20 nm (e) and $d$=210 nm (f) for $F=DT_d^{SP}$ at different time delays $t_d$.

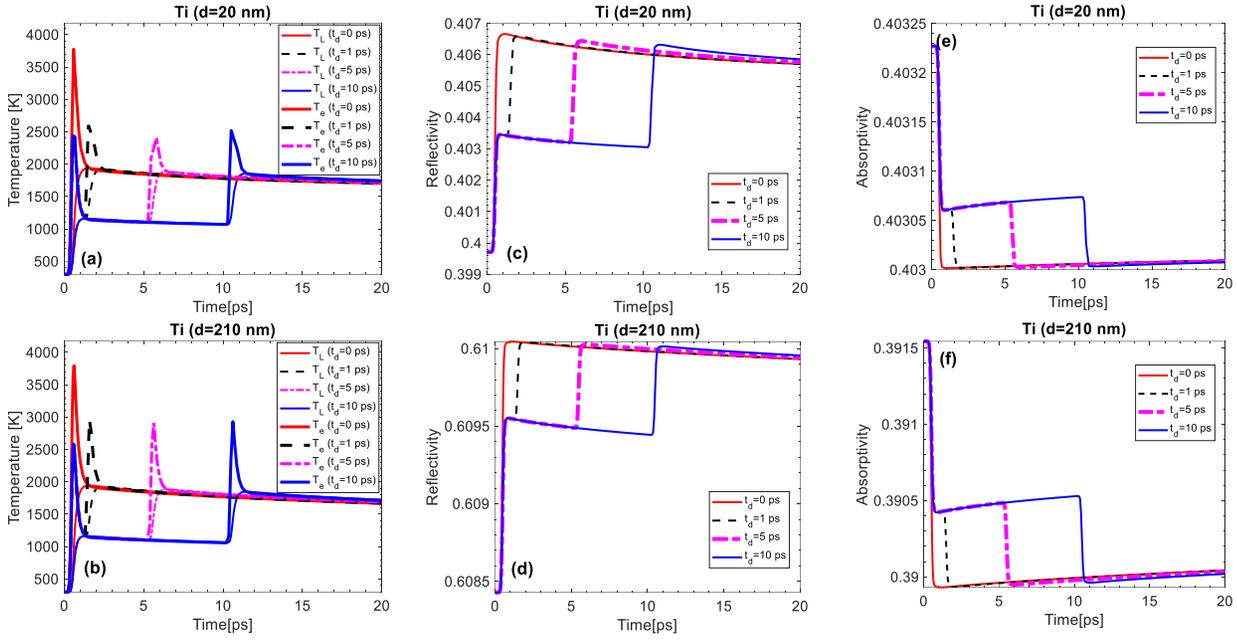

**Figure SM10:** Results for Ti at different pulse delays and material thicknesses: Electron and Lattice Temperature evolution at $d$=20 nm (a) and $d$=210 nm (b) for $F=DT_d^{SP}$ at different time delays $t_d$; Reflectivity evolution at $d$=20 nm (c) and $d$=210 nm (d) for $F=DT_d^{SP}$ at different time delays $t_d$ (c); Absorptivity evolution at $d$=20 nm (e) and $d$=210 nm (f) for $F=DT_d^{SP}$ at different time delays $t_d$.

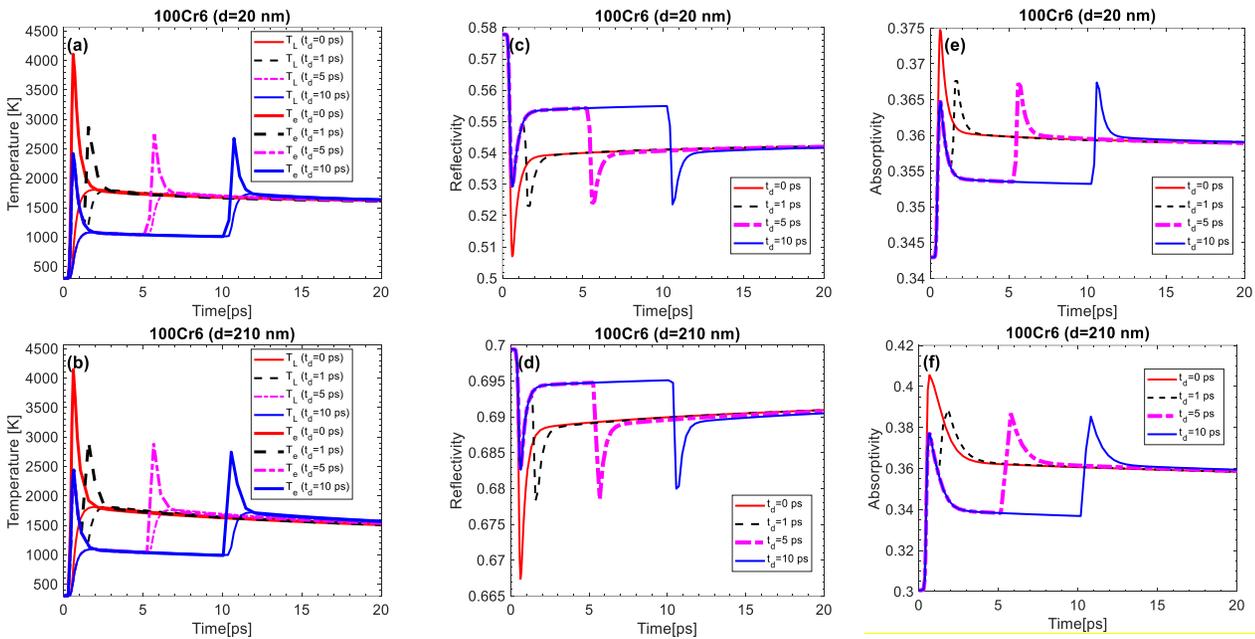

**Figure SM11:** Results for 100Cr6 at different pulse delays and material thicknesses: Electron and Lattice Temperature evolution at $d$=20 nm (a) and $d$=210 nm (b) for $F=DT_d^{SP}$ at different time delays $t_d$; Reflectivity evolution at $d$=20 nm (c) and $d$=210 nm (d) for $F=DT_d^{SP}$ at different time delays $t_d$ (c); Absorptivity evolution at $d$=20 nm (e) and $d$=210 nm (f) for $F=DT_d^{SP}$ at different time delays $t_d$.

## 2. Numerical procedure to calculate LIDT

The damage threshold is determined through an iterative, sign-controlled numerical procedure coupled with the Two-Temperature Model assuming the evolution of the optical parameters (Eqs.1-3). A trial value of the laser fluence is defined and the coupled electron–lattice temperature equations are solved numerically. From the solution, the resulting lattice temperature is evaluated and compared to the critical melting temperature. At each iteration, the relative error between the calculated temperature and the melting temperature is computed. If this relative error falls below a prescribed tolerance (i.e. 0.2% in this study), the current trial value of the fluence is identified as the damage threshold and the iteration is terminated. If convergence is not achieved, the signed difference between the calculated temperature and the melting temperature is evaluated. The sign of this difference indicates whether the current trial value overestimates or underestimates the damage threshold. When a change in sign is detected between two consecutive iterations, indicating that the threshold has been crossed, the step size used to update the fluence is reduced by a factor of two in order to improve numerical stability and accuracy. The new trial fluence value is then updated by increasing or decreasing its value depending on the sign of the current error: if the calculated temperature exceeds the melting temperature, the fluence is reduced; otherwise, it is increased. The signed error from the current iteration is stored and used in the subsequent step to monitor sign changes. This iterative process continues until the relative error between the calculated lattice temperature and the melting temperature satisfies the prescribed tolerance. The resulting fluence is taken as the damage threshold.

The above procedure is used to generate Figures 1-11 in the main manuscript for all values of the pulse separation and film thickness.